\documentclass[twocolumn]{aastex631}

\usepackage{dsnr}
\usepackage[normalem]{ulem}

\received{11 Dec 2023}
\revised{21 Mar 2024}
\accepted{29 Mar 2024}

\submitjournal{ApJ}
\shorttitle{ORC Spectra}
\shortauthors{Rupke et al.}

\begin{document}

\title{The Intrinsic Sizes of Odd Radio Circles}

\correspondingauthor{David Rupke}
\email{drupke@gmail.com}

\author[0000-0002-1608-7564]{David S. N. Rupke}
\affiliation{Department of Physics, Rhodes College, 2000 North Parkway, Memphis, TN 38112, USA}
\affil{Zentrum für Astronomie der Universität Heidelberg, Astronomisches Rechen-Institut, Mönchhofstr 12-14, D-69120 Heidelberg, Germany}

\author[0000-0002-2583-5894]{Alison L. Coil}
\affil{Department of Astronomy and Astrophysics, University of California, San Diego, La Jolla, CA 92093, USA}

\author[0000-0002-8571-9801]{Kelly E. Whalen}
\affil{Department of Physics and Astronomy, Dartmouth College, Hanover, NH 03755, USA}
\affil{Goddard Space Flight Center, Greenbelt, MD}

\author[0000-0002-2733-4559]{John Moustakas}
\affil{Department of Physics and Astronomy, Siena College, Loudonville, NY 12211, USA}

\author[0000-0003-3097-5178]{Christy A. Tremonti}
\affil{Department of Astronomy, University of Wisconsin-Madison, Madison, WI 53706, USA}

\author[0000-0002-2451-9160]{Serena Perrotta}
\affil{Department of Astronomy and Astrophysics, University of California, San Diego, La Jolla, CA 92093, USA}

\begin{abstract}
A new class of radio source, the so-called Odd Radio Circles (ORCs), have been discovered by recent sensitive, large-area radio continuum surveys. The distances of these sources have so far relied on photometric redshifts of optical galaxies found at the centers of or near the ORCs. Here we present Gemini rest-frame optical spectroscopy of six galaxies at the centers of, or potentially associated with, the first five ORC discoveries. We supplement this with Legacy Survey imaging and Prospector fits to their $griz$$+${\em W1/W2} photometry. Of the three ORCs with central galaxies, all lie at distances ($z = 0.27-0.55$) that confirm the large intrinsic diameters of the radio circles (300--500~kpc). The central galaxies are massive ($M_*\sim10^{11}$~\msun), red, unobscured ellipticals with old ($\ga$1~Gyr) stellar populations. They have LINER spectral types that are shock- or AGN-powered. All three host low-luminosity, radio-quiet AGN. The similarity of their central galaxies are consistent with a common origin, perhaps as a blastwave from an ancient starburst. The other two ORCs are adjacent and have no prominent central galaxies. However, the $z=0.25$ disk galaxy that lies between them hosts a Type 2, moderate-luminosity AGN. They may instead be the lobes of a radio jet from this AGN.
\end{abstract}

%\keywords{Circumgalactic medium (1879), Stellar feedback (1602), Galactic winds (572), Shocks (2086)}

\section{Introduction} \label{sec:intro}

The Odd Radio Circles (ORCs) are a new class of radio source \citep{2021PASA...38....3N}. ORCs are large rings of faint, diffuse continuum emission spanning $\sim$1 arcminute. The first examples of this class were detected at 1~GHz in the Evolutionary Map of the Universe (EMU; \citealt{2011PASA...28..215N}) Pilot Survey \citep{2021PASA...38....9H} using the Australian Square Kilometer Array Pathfinder (ASKAP). The EMU Pilot Survey is able to find rare, faint objects because it achieves $\sim$12\arcsec\ resolution; is sensitive to low surface brightness emission, reaching $\sim$30~$\mu$Jy~beam$^{-1}$; and covers 270 deg$^2$ of the sky.

Several other ORCs, or ORC-like structures, have subsequently been discovered. \citet{2021PASA...38....3N} present ORC4 alongside ORCs 1--3 after finding it in reprocessed 325~MHz Giant MeterWave Radio Telescope (GMRT) data \citep{2017A&A...603A.125V,2017A&A...598A..78I}.  \citet{2021MNRAS.505L..11K} discovered ORC J0102$-$2450 (labeled ORC5 by \citealt{2021Galax...9...83N}) in later ASKAP data. Inspired by previous detections, \citet{2022RNAAS...6..100O} describe an ORC in Low Frequency Array (LOFAR) survey data. \citet{2022MNRAS.512..265F} show an exceptionally large example, J0624$-$6948, observed with ASKAP. \citet{2022PASA...39...51G} deliver two ORC "candidates" and \citet{2023MNRAS.520.1439L} describe an ORC-like structure; all three were found using machine learning techniques applied to low-frequency survey data (ASKAP and MeerKAT). 

%Most recently, \citet{2023arXiv230411784K} detected ORC J1027$–$4422 in deep 1.3 GHz MeerKAT imaging.

Nine of these ten structures are probably extragalactic in origin. The most likely Galactic origin of ORCs, that of a disk supernova remnant, is largely ruled out by the high Galactic latitudes of most ORCs or ORC candidates: $|b|\ga20^\circ$ \citep{2021PASA...38....3N,2022arXiv220910554S}. The exception is J0624$-$6948, which is 3$^\circ$ from the radio edge of the Large Magellanic Cloud. Circumgalactic or intragroup supernova remnants are also possible but unlikely for the other ORCs \citep{2022arXiv220910554S,2022MNRAS.513.1300N}

%The exceptions are ORC~J1027$-$4422, which has $b=11\fdg3$ \citep{2023arXiv230411784K}, and 

The remaining nine sources consist of large, limb-brightened circular objects that often have clumpy substructure in the rings. Many also contain irregular interior emission. At least 5/9 have a galaxy at or near the geometric center of the ORC with a photometric redshift in the range $z=0.27-0.56$: ORCs 1, 4, and 5 \citep{2022MNRAS.513.1300N}; EMU-PS J222339.5$-$483449 \citep{2022PASA...39...51G}; and the so-called SAURON object \citep{2023MNRAS.520.1439L}. Adopting these redshifts, the intrinsic sizes of the ORCs are several hundred kpc in diameter.

The physical mechanism that produces such enormous radio structures is yet unknown. Candidates include a blast wave from an explosive event due to star formation or active galactic nuclei \citep{2022MNRAS.513.1300N,2024Natur.625..459C}, a shockwave from an extreme galaxy merger \citep{2023ApJ...945...74D}, or a galaxy virial shock \citep{2024MNRAS.528.3854Y}.

Higher-resolution, multi-band radio measurements of ORC1 are consistent with the blastwave scenario \citep{2022MNRAS.513.1300N}. In \citet{2024Natur.625..459C}, we advance further evidence in favor of this model through deep, rest-frame, near-UV/optical integral field spectroscopy of ORC4 with the Keck Cosmic Web Imager (KCWI). These KCWI data constrain the redshift of the central source to $z=0.4512$ and reveal strong [\ion{O}{2}]~3727,~3729~\AA\ ionized gas emission. This emission extends to 20 kpc radius, shows a velocity gradient and high velocity dispersion (150--200~\kms), and has very high equivalent width (50~\AA). Photometry of the 10$^{11}$~\msun\ central galaxy is consistent with a 1~Gyr-old burst of star formation that formed half of the galaxy's stars. Accompanying numerical simulations suggest that this burst could have produced the large-scale radio emission through a forward shock driven by a powerful wind. The 10$\times$ smaller ionized gas emission is the result of gas heated by the reverse shock that then falls back towards the central galaxy and produces further, small-scale shocks.

To explore the nature of the central galaxies of ORCs, we here present new rest-frame optical spectroscopy ($\sim$3700 to 6800~\AA) of the central galaxies of ORCs 1, 4, and 5. We also target several galaxies near these central galaxies. For ORCs 2/3, which are spatially adjacent but do not contain prominent central galaxies, we target a bright galaxy located in projection between their centers. We also analyze deep photometry from the Dark Energy Spectroscopic Instrument (DESI) Legacy Surveys to constrain the masses and star formation history of the associated galaxies.   

In Section~\ref{sec:obs-red}, we present the spectroscopic observations and data reduction, multicolor images of the ORC fields, and the Tractor photometry. In Section~\ref{sec:results}, we present the resulting spectra, fits to the spectra and photometry, and the resulting physical quantities derived from these fits. We discuss the properties of the galaxies in light of these data, as well as the implications for the origin of ORCs, in Section~\ref{sec:discuss}. We conclude in Section~\ref{sec:conclude}. In all calculations, we assume a flat $\Lambda$ cosmology with $\Omega_m=0.315$ and $H_0=67.4$~\kms~Mpc$^{-1}$. In our star formation rate (SFR) calculations, we assume a Salpeter IMF.

\section{Observations and Data Reduction} \label{sec:obs-red}
\subsection{Spectroscopy} \label{sec:spec}
\setlength\tabcolsep{2pt}
\begin{deluxetable}{lcrccr}
  \tablecaption{Observations\label{tab:obs}}
  \tabletypesize{\footnotesize}
  \tablewidth{0pt}

  \tablehead{\colhead{Target} & \colhead{Instrument} & \colhead{Date} & \colhead{$t_\mathrm{exp}$} & \colhead{$\lambda_\mathrm{cen}$} & \colhead{PA} \\
  \colhead{} & \colhead{} & \colhead{} & \colhead{s} & \colhead{nm} & \colhead{$\circ$}}
  \colnumbers
  \startdata
    ORC1   & GMOS-S & 30 Aug 2022 & 941$\times$3 & 705/750/795 &  83.8 \\
    ORC2/3 & GMOS-S & 19 Jun 2022 & 940$\times$3 & 705/750/795 & 270.0 \\
    ORC4   & GMOS-N &  5 Jun 2022 & 704$\times$4 & 710/714     & 153.0 \\
    ORC5   & GMOS-S & 11 Jun 2022 & 941$\times$3 & 705/750/795 & 315.8 \\
  \enddata

  \tablecomments{Column 1: Science target. Column 2: Instrument. Column 3: UTC observation date. Column 4: Exposure time, in s. Column 5: Central wavelength of grating tilt, in nm. Column 6: Sky position angle of slit, in degrees East of North.}
\end{deluxetable}
\begin{figure*}
    \centering
    \includegraphics[width=\textwidth]{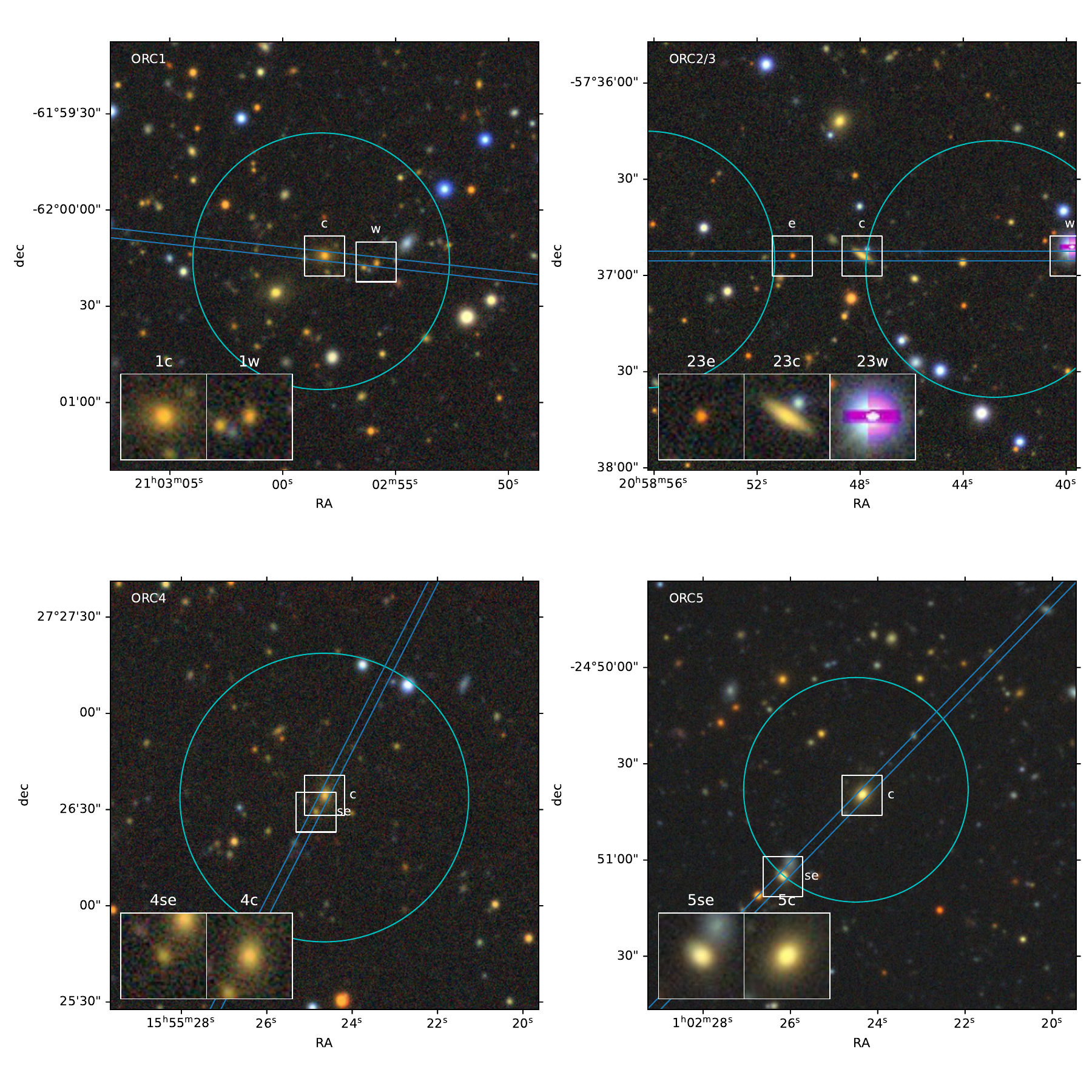}
    \caption{DESI Legacy Imaging Survey DR10 $griz$ color images of the target fields. Sizes are approximately $2\arcmin\times2\arcmin$, which is roughly double the diameter of the ORCs \citep{2021PASA...38....3N,2021MNRAS.505L..11K}. Cyan circles represent the ORC extents, with centers from \citet{2021PASA...38....3N} and \citet{2021MNRAS.505L..11K} and diameters from \citet{2022MNRAS.513.1300N}. Slits are overlaid, enlarged in width for presentation. Extracted sources are labeled with boxes, and zoomed-in cutouts are in the lower left of each panel.}
    \label{fig:img-slit}
\end{figure*}
\begin{deluxetable*}{lcccccr}
  \tablecaption{Coordinates and Morphology\label{tab:coord-shape}}
  \tablewidth{0pt}
  \tablehead{\colhead{Label} & \colhead{RA} & \colhead{dec} & \colhead{TYPE} & \colhead{$n$} & \colhead{$R_e$} & \colhead{$f_{p/s}$} \\   
  \colhead{} & \colhead{hms} & \colhead{dms} & \colhead{}  & \colhead{} & \colhead{\arcsec}}
  \colnumbers
  \startdata
ORC1c  &  21:02:58.15$\pm$0.12 & $-$62:00:14.404$\pm$0.004 & SER &    6.0 &   3.52 &    4.05\\
ORC23c &  20:58:47.92$\pm$0.08 & $-$57:36:53.944$\pm$0.002 & EXP &    1.0 &   1.69 &    1.96\\
ORC4c  &  15:55:24.64$\pm$0.09 & $+$27:26:34.311$\pm$0.007 & DEV &    4.0 &   2.76 &    3.50\\
ORC4se &  15:55:24.85$\pm$0.23 & $+$27:26:29.230$\pm$0.016 & DEV &    4.0 &   0.55 &    1.85\\
ORC5c  &  01:02:24.35$\pm$0.08 & $-$24:50:39.736$\pm$0.004 & SER &    5.4 &   1.07 &    1.69\\
ORC5se &  01:02:26.17$\pm$0.80 & $-$24:51:05.100$\pm$0.025 & SER &    3.3 &   0.86 &    1.98\\
\hline
ORC1w  &  21:02:55.85$\pm$0.20 & $-$62:00:16.799$\pm$0.008 & DEV &    4.0 &   0.44 &    3.85\\
ORC23e &  20:58:50.62$\pm$0.14 & $-$57:36:54.033$\pm$0.005 & PSF &\nodata &\nodata &    1.12\\
ORC23w &  20:58:39.79$\pm$0.00 & $-$57:36:51.256$\pm$0.000 & PSF &\nodata &\nodata &  142.40\tablenotemark{a}
  \enddata

  \tablecomments{Column 1: Label from Figure~\ref{fig:img-slit}. Columns 2--3: Coordinates from DESI Legacy Imaging Surveys. Column 4: The fitted morphological type (PSF $=$ point source/star; SER $=$ S\'{e}rsic profile; EXP $=$ exponential profile (spiral galaxy); and DEV $=$ deVaucoulers profile (elliptical galaxy). Column 5: S\'{e}rsic index. Fixed for EXP or DEV types; allowed to vary for SER. Column 6: Half-light (effective) radius, in arcseconds. Column 7: Average ratio in $r$ and $i$ bands between DESI photometry and synthetic photometry from spectra.}
\tablenotetext{a}{This value is unusually high because the slit only grazes the star's point spread function.}  
\end{deluxetable*}
\begin{deluxetable*}{lcccccccc}
  \tablecaption{Photometry\label{tab:phot}}
  \tablewidth{0pt}
  \tabletypesize{\scriptsize}
  \tablehead{\colhead{Label} & \colhead{$g$} & \colhead{$r$} & \colhead{$i$} & \colhead{$z$} & \colhead{$W1$} & \colhead{$W2$} & \colhead{$W3$} & \colhead{$W4$}}
  \colnumbers

  \startdata
ORC1c  & 21.485$^{+0.031}_{-0.030}$ & 19.544$^{+0.006}_{-0.006}$ & 18.668$^{+0.007}_{-0.007}$ & 18.221$^{+0.007}_{-0.007}$ & 17.234$^{+0.008}_{-0.008}$ & 17.710$^{+0.023}_{-0.022}$ &              $>$17.467 &              $>$14.248 \\
ORC23c & 20.761$^{+0.008}_{-0.008}$ & 19.473$^{+0.004}_{-0.004}$ & 18.946$^{+0.005}_{-0.005}$ & 18.629$^{+0.004}_{-0.004}$ & 18.139$^{+0.012}_{-0.012}$ & 18.277$^{+0.028}_{-0.027}$ & 16.735$^{+0.190}_{-0.162}$ &              $>$14.424 \\
ORC4c  & 21.028$^{+0.019}_{-0.018}$ & 19.524$^{+0.007}_{-0.007}$ & 18.910$^{+0.008}_{-0.008}$ & 18.537$^{+0.010}_{-0.010}$ & 17.891$^{+0.012}_{-0.011}$ & 18.372$^{+0.035}_{-0.033}$ &              $>$17.922 &              $>$14.893 \\
ORC4se & 23.559$^{+0.095}_{-0.087}$ & 21.809$^{+0.027}_{-0.026}$ & 21.082$^{+0.027}_{-0.027}$ & 20.786$^{+0.038}_{-0.037}$ & 20.241$^{+0.070}_{-0.066}$ & 20.682$^{+0.235}_{-0.193}$ &              $>$18.156 &              $>$15.019 \\
ORC5c  & 20.247$^{+0.005}_{-0.005}$ & 18.788$^{+0.002}_{-0.002}$ & 18.309$^{+0.001}_{-0.001}$ & 18.032$^{+0.001}_{-0.001}$ & 17.894$^{+0.010}_{-0.010}$ & 18.344$^{+0.031}_{-0.030}$ &              $>$17.401 & 14.077$^{+0.335}_{-0.256}$ \\
ORC5se & 20.355$^{+0.004}_{-0.004}$ & 19.338$^{+0.003}_{-0.003}$ & 18.806$^{+0.001}_{-0.001}$ & 18.564$^{+0.002}_{-0.002}$ & 18.175$^{+0.013}_{-0.013}$ & 18.159$^{+0.026}_{-0.026}$ & 16.382$^{+0.148}_{-0.130}$ &              $>$14.357 \\
\hline
ORC1w  & 23.518$^{+0.085}_{-0.079}$ & 21.601$^{+0.014}_{-0.014}$ & 20.770$^{+0.024}_{-0.024}$ & 20.335$^{+0.017}_{-0.017}$ & 19.138$^{+0.027}_{-0.026}$ & 19.773$^{+0.100}_{-0.092}$ &              $>$17.801 &              $>$14.442 \\
ORC23e & 24.350$^{+0.115}_{-0.104}$ & 22.704$^{+0.036}_{-0.035}$ & 21.128$^{+0.019}_{-0.018}$ & 20.476$^{+0.012}_{-0.012}$ & 20.657$^{+0.096}_{-0.088}$ &              $>$21.263 &              $>$17.590 &              $>$14.465 \\
ORC23w & 15.073$^{+0.000}_{-0.000}$ & 14.717$^{+0.001}_{-0.001}$ & 14.583$^{+0.001}_{-0.001}$ & 14.588$^{+0.000}_{-0.000}$ & 16.031$^{+0.003}_{-0.003}$ & 16.712$^{+0.008}_{-0.008}$ &              $>$17.593 &              $>$14.450
  \enddata
  \tablecomments{Column 1: Label from Figure~\ref{fig:img-slit}. Columns 2--9: Tractor \citep{2016ascl.soft04008L,2016AJ....151...36L} photometry from DESI Legacy Surveys \citep{2019AJ....157..168D}. Magnitudes are AB and calculated directly from the catalog flux in nanomaggies using $m = 22.5 - 2.5\,\mathrm{log}_{10}(\mathrm{flux})$. We have also corrected each flux for Galactic extinction by dividing by MW\_TRANSMISSION from the Legacy Survey catalog. We require 3$\sigma$ detections, and otherwise list 3$\sigma$ lower limits in AB mags.}
\end{deluxetable*}
We observed ORCs 1--5 with the Gemini Multi-Object Spectrographs on Gemini North and South (GMOS-N and GMOS-S; \citealt{2004PASP..116..425H, 2016SPIE.9908E..2SG}). In designing the observations, we specified observing conditions of 70th-percentile image quality (IQ70) and 50th-percentile cloud cover (CC50). The actual observations are summarized in Table~\ref{tab:obs}. We used the R400 grating with a 0\farcs75 slit to achieve a spectral resolution of $\delta\lambda = 4.0$~\AA. The GG455 filter reduces second-order contamination. We binned 2$\times$2 to increase sensitivity. To minimize the effects of detector artifacts and span chip gaps, we dithered in wavelength. We especially needed to mitigate the bad amplifier 5 of CCD2 on GMOS-S, but also some enhanced noise elsewhere in the GMOS-S detector and a hot column in amplifier 5 of GMOS-N. We also dithered along the slit by $\pm$1\arcsec. The resulting observed wavelength ranges are 4770--9560~\AA\ and 4730--10340~\AA\ for GMOS-N and GMOS-S, respectively.

In Figure~\ref{fig:img-slit} we show images of the ORC fields (see Section~\ref{sec:imgs}). For ORCs 1, 4, and 5, we chose the slit position and PA to cover the galaxy at the center of the ORC and one other source within the ORC footprint. Following the lead of \citet{2022MNRAS.513.1300N}, we denote the central sources as ORC1c, ORC4c, and ORC5c. %(\citealt{2021PASA...38....3N} refer to ORC4C as ORC4G, and \citealt{2022MNRAS.513.1300N} do not label the central source.)
For ORCs 2$+$3, we centered the slit on the galaxy directly between the two ORCs (also called ORC2/3c) and chose a PA crossing the center of each ORC. % and overlapping sources seen in imaging.

We observed the flux standards BD$+$28~4211 and LTT~7379 for calibration of the GMOS-N and -S data, respectively. For the GMOS-N data, the flux standard was observed only at central wavelengths 705 and 795~nm. For GMOS-S, the standard was observed at the mean of the central wavelengths, 712~nm. However, second-order contamination is a problem above 9000~\AA\ for the GMOS-N standard. We found in the Gemini Observatory Archive data of EG~131 taken in 2022 Sep. with the same configuration as BD$+$28~4211, except using the RG610 filter. We applied the resulting sensitivity function from this standard above 9000~\AA, but used the original sensitivity function for lower wavelengths.

We reduced the data using v1.11 and v1.12 of {\ttfamily PypeIt} \citep{pypeit:joss_pub}. We retrieved from the Gemini Observatory Archive individual bias frames taken within a few days of each set of observations for bias subtraction. By studying the GCAL flats, we observed changes in the positions of bad/unexposed areas of the detector (bad columns, amplifiers, or slit gaps) compared to those specified by the default bad pixel masks (bpms) in {\ttfamily PypeIt}. This includes changes in the slit gaps in GMOS-S between the June and August observations when a detector repair was attempted. To account for this, we edited by hand the default bpm for each configuration. We also encountered large RMS values when applying the default wavelength calibrations to the observed CuAr lamp exposures. To mitigate this, we ran {\ttfamily pypeit\_identify} with a curated linelist on one of the extracted spectra in each wavelength configuration. {\ttfamily PypeIt} also converts to vacuum wavelengths and corrects to heliocentric velocity. 

During flux calibration, we used the IR algorithm for computing the sensitivity function. We corrected the data for airmass differences with the flux standards where necessary. We did not, however, apply slit loss corrections arising from not observing at the parallactic angle. For our data, we estimate these wavaelength-dependent losses to be small ($\la10\%$ correction at the edges of the data\footnote{\url{https://www.gemini.edu/observing/resources/optical-resources\#ADR}}) because either the difference from parallactic was $<20^\circ$ and/or the airmass was $\leq1.2$ and because the target was acquired in the $r$ band. We also applied a correction for Galactic extinction to each spectrum \citep{2011ApJ...737..103S}.

\subsection{Images} \label{sec:imgs}

We retrieve coordinates and morphology (Table~\ref{tab:coord-shape}), photometry (Table~\ref{tab:phot}), and $griz$ images (Figure~\ref{fig:img-slit}) from Data Release 10.1 of the DESI Legacy Imaging Surveys \citep{2019AJ....157..168D}. The shape information and photometry are derived from Tractor\footnote{\url{https://github.com/dstndstn/tractor}} \citep{2016ascl.soft04008L,2016AJ....151...36L} models of each source. For further details on the imaging and shape modeling, see \citet{2023ApJS..269....3M}.
%\end{rotatetable}
To verify our spectrophotometry, we create synthetic $r$ and $i$ fluxes with {\ttfamily pyphot} \citep{Fouesneau_pyphot_2022} and the DESI Imaging Survey ($grz$) and DECam DR1 ($i$) response curves tabulated in {\ttfamily speclite}\footnote{\url{https://speclite.readthedocs.org}}. The average ratio between the Tractor fluxes (corrected for Galactic extinction) and our synthetic photometry are listed in Table~\ref{tab:coord-shape} as $f_{p/s}$. For the small M-dwarf covered completely by one of our slits (ORC23e), this ratio is close to unity. (The other star is only barely in the slit, resulting in $f_{p/s} > 100$.) For the other 7 sources that are spatially-resolved in imaging, this ratio ranges from 1.7--4. We attribute this mainly to slit losses. This is borne out by the fact that the four sources with modeled half-light radius $R_\mathrm{e}<2\arcsec$ have $f_{p/s} = 1.7-2.0$, while the larger sources have $f_{p/s}=3.5-4.1$. (ORC1w does not fit this pattern because it does not fall completely in the slit). However, we also caution that the Tractor photometry is based on images taken in seeing sometimes as large as FWHM$\sim$1\farcs7. KCWI $g$-band integral field data of ORC4c give $R_\mathrm{e}=0\farcs8$ \citep{2024Natur.625..459C}, and the flux ratio of the KCWI-to-GMOS spectra over identical wavelength ranges is $\sim$1.5. This compares to $R_\mathrm{e}=2\farcs8$ and $f_{p/s} = 3.5$ from the Tractor catalog.

Regardless of the normalization, the shape of the source SEDs from Tractor photometry match well with our spectrophotometry. This is apparent visually in the overlap of the (scaled) photometry with our spectra in Figures~\ref{fig:spec}--\ref{fig:spec-unided}.

\section{Results and Modeling} \label{sec:results}

\subsection{Spectra} \label{sec:results-spec}

We extracted nine spectra from the four fields based on their continuum traces. We visually inspected the sky-subtracted 2D data for extended line emission along the slit, but none was apparent. Six of these are galaxies for which we determined redshifts and other properties from spectral fits (Figure~\ref{fig:spec}). For three others (Figure~\ref{fig:spec-unided} in Appendix~\ref{sec:appendix}), two are likely stars (ORC23e and ORC23w). The spectrum of ORC23e resembles an M-dwarf, and ORC23w a hotter, bluer star. The photometry confirms this, as we discuss above. The final spectrum, of ORC1w, is low signal-to-noise in part because the object did not fully fall in the slit; we do not identify the source.
\begin{figure*}
    \centering
    \includegraphics[width=\textwidth]{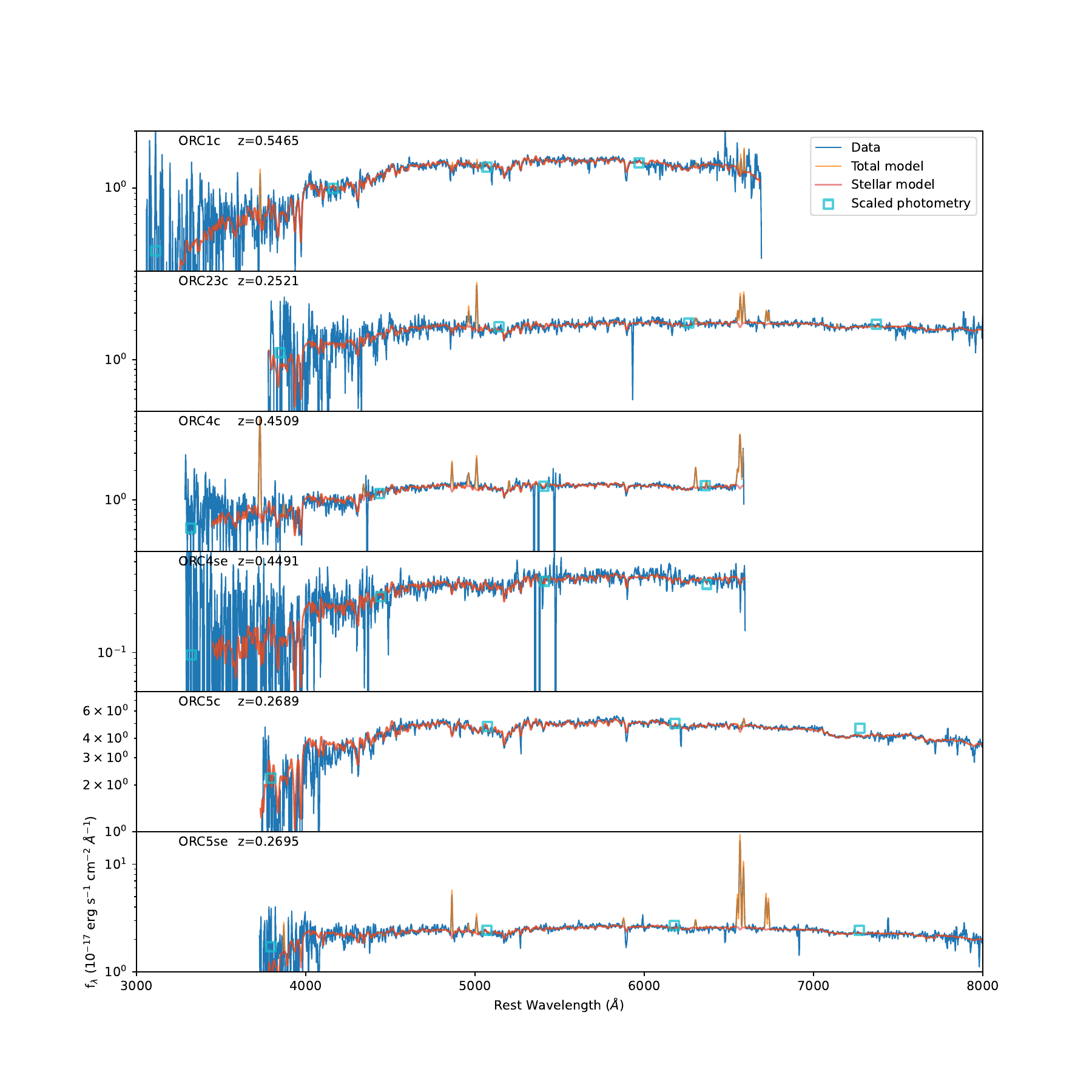}
    \caption{Rest-frame spectra of the six galaxies with identifiable redshifts. The data (blue) are smoothed with a 5-pixel boxcar; the continuum model (red) and emission-line model (orange) are overplotted. Photometry (cyan boxes) is shown, scaled down by the mean $r+i$ slit loss $f_{p/s}$ (Column 7 of Table~\ref{tab:coord-shape}).}
    \label{fig:spec}
\end{figure*}
\begin{figure*}
    \centering
    \includegraphics[width=\textwidth]{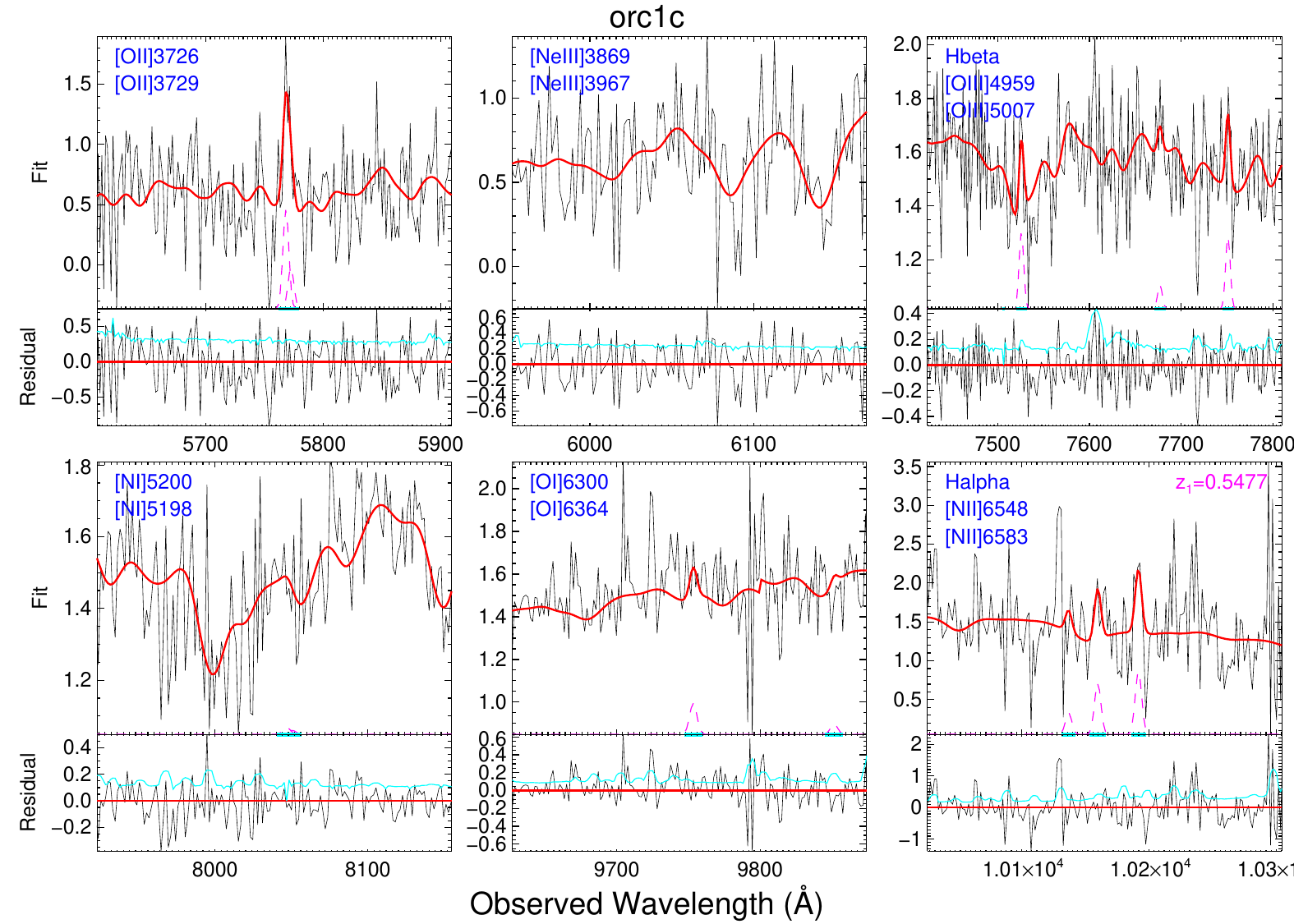}
    \caption{Spectra of galaxies with detectable line emission, displayed to highlight individual lines. The upper portion of each panel (``Fit'') shows the data (black); the best-fit model (red); the emission line model (dashed magenta); and the wavelength range masked during continuum fitting (cyan bars). The lower portion (``Residual'') shows the data minus the model, in addition to 1$\sigma$ error spectra (cyan). Lines that may appear in the wavelength range of each panel are listed in the upper left corner of that panel.}
    \label{fig:spec-lines}
\end{figure*}
\setcounter{figure}{2}
\begin{figure*}
    \centering
    \includegraphics[width=\textwidth]{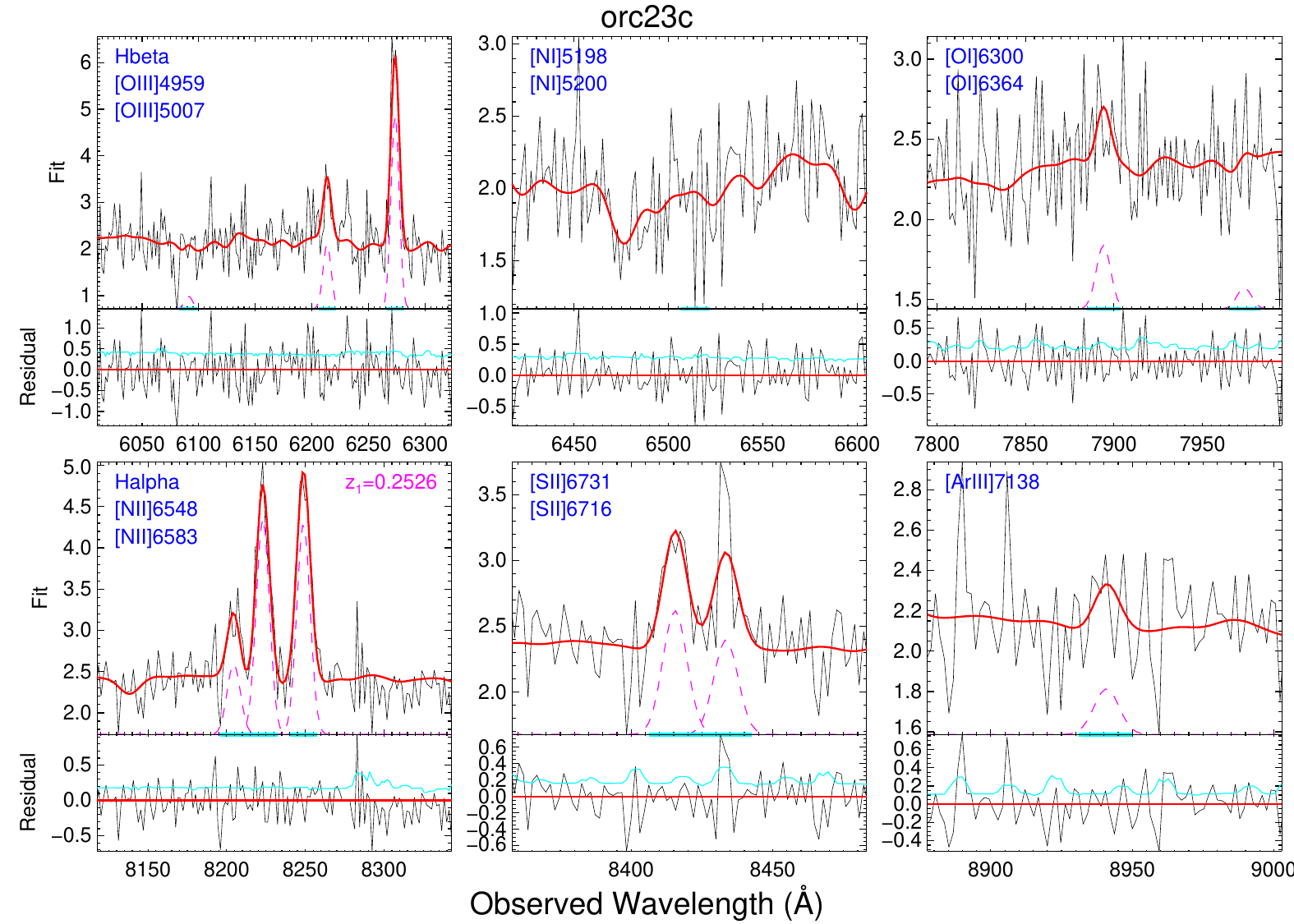}
    \caption{{\it Continued.}}
\end{figure*}
\setcounter{figure}{2}
\begin{figure*}
    \centering
    \includegraphics[width=\textwidth]{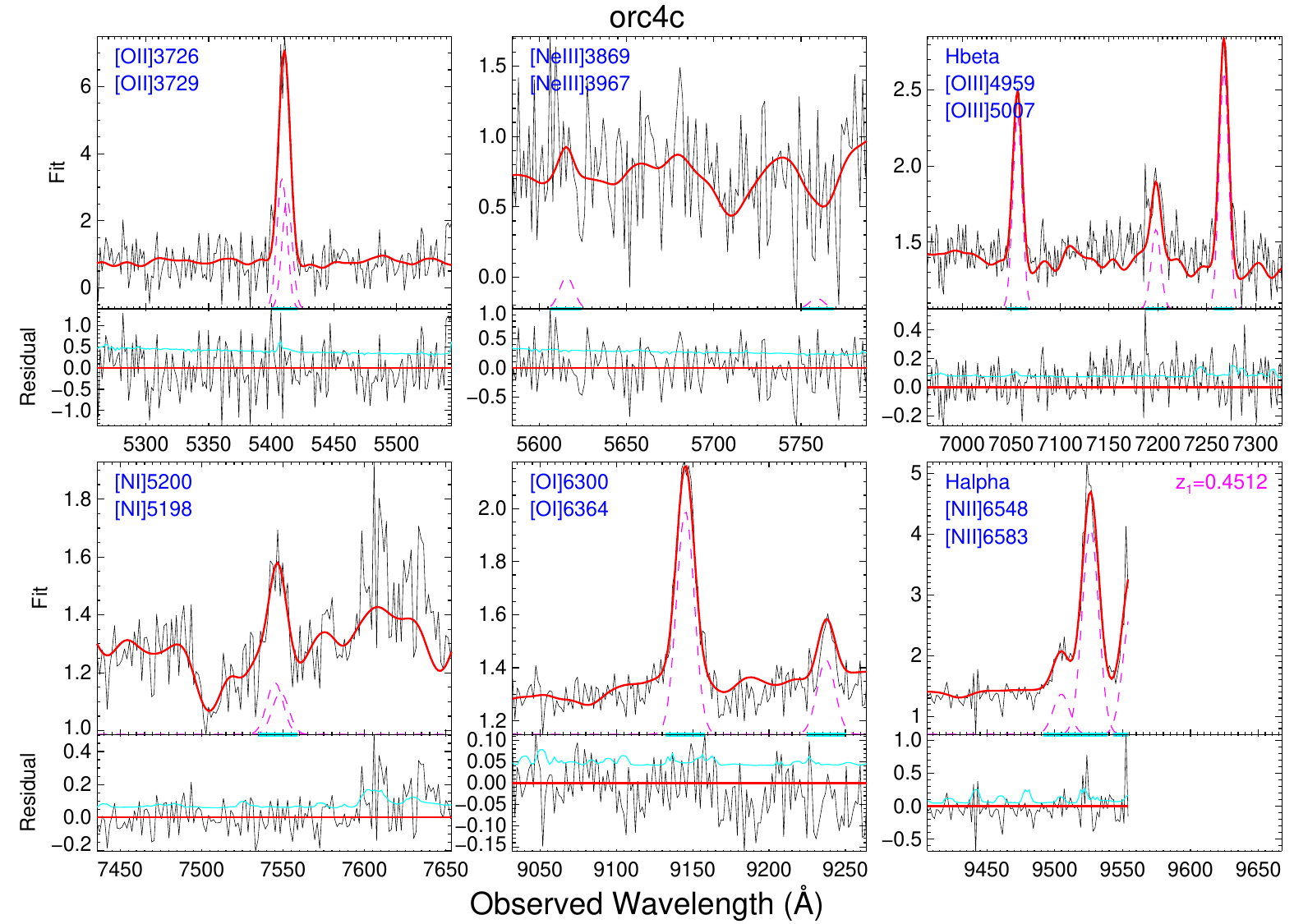}
    \caption{{\it Continued.}}
\end{figure*}
\setcounter{figure}{2}
\begin{figure*}
    \centering
    \includegraphics[width=\textwidth]{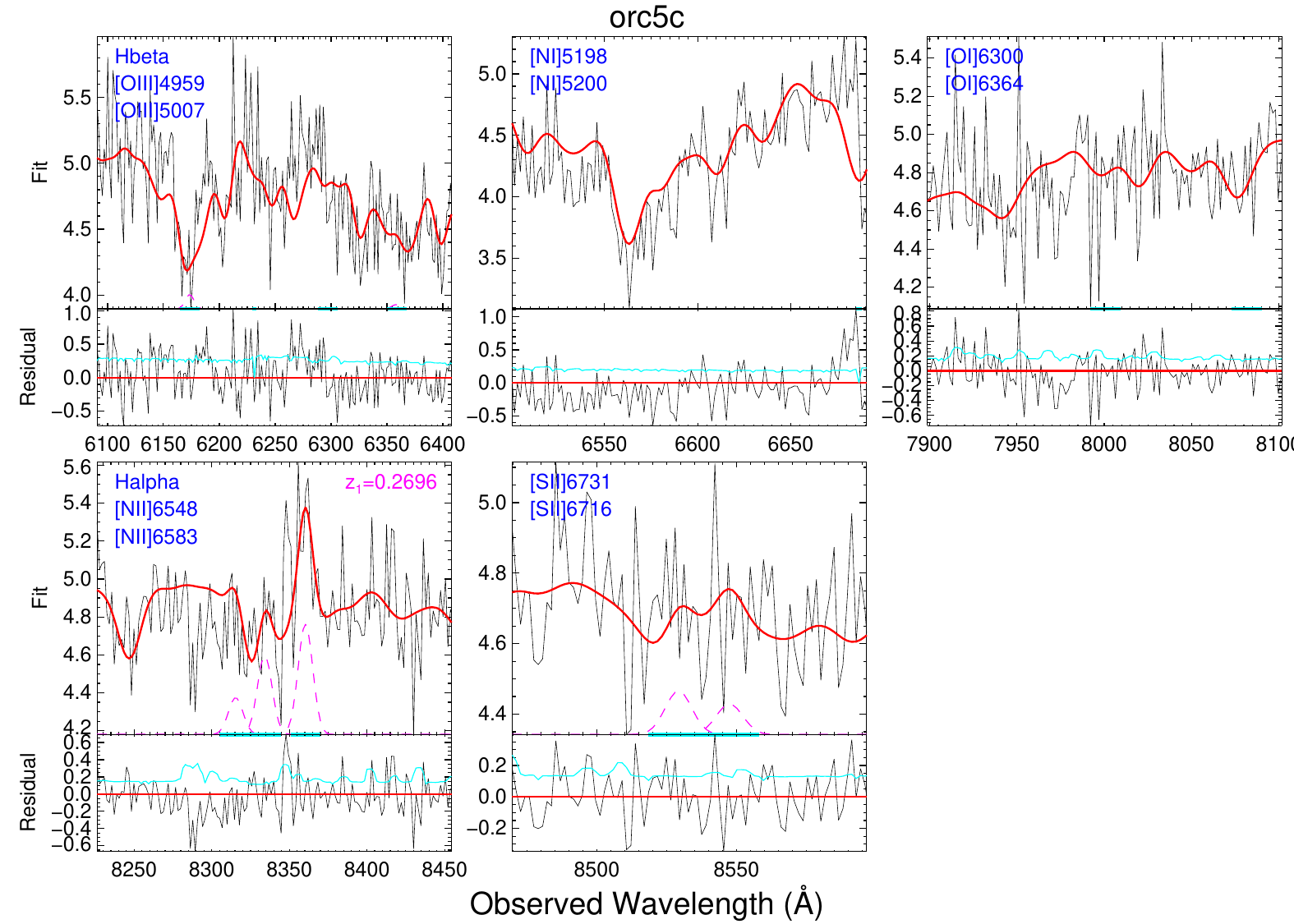}
    \caption{{\it Continued.}}
\end{figure*}
\setcounter{figure}{2}
\begin{figure*}
    \centering
    \includegraphics[width=\textwidth]{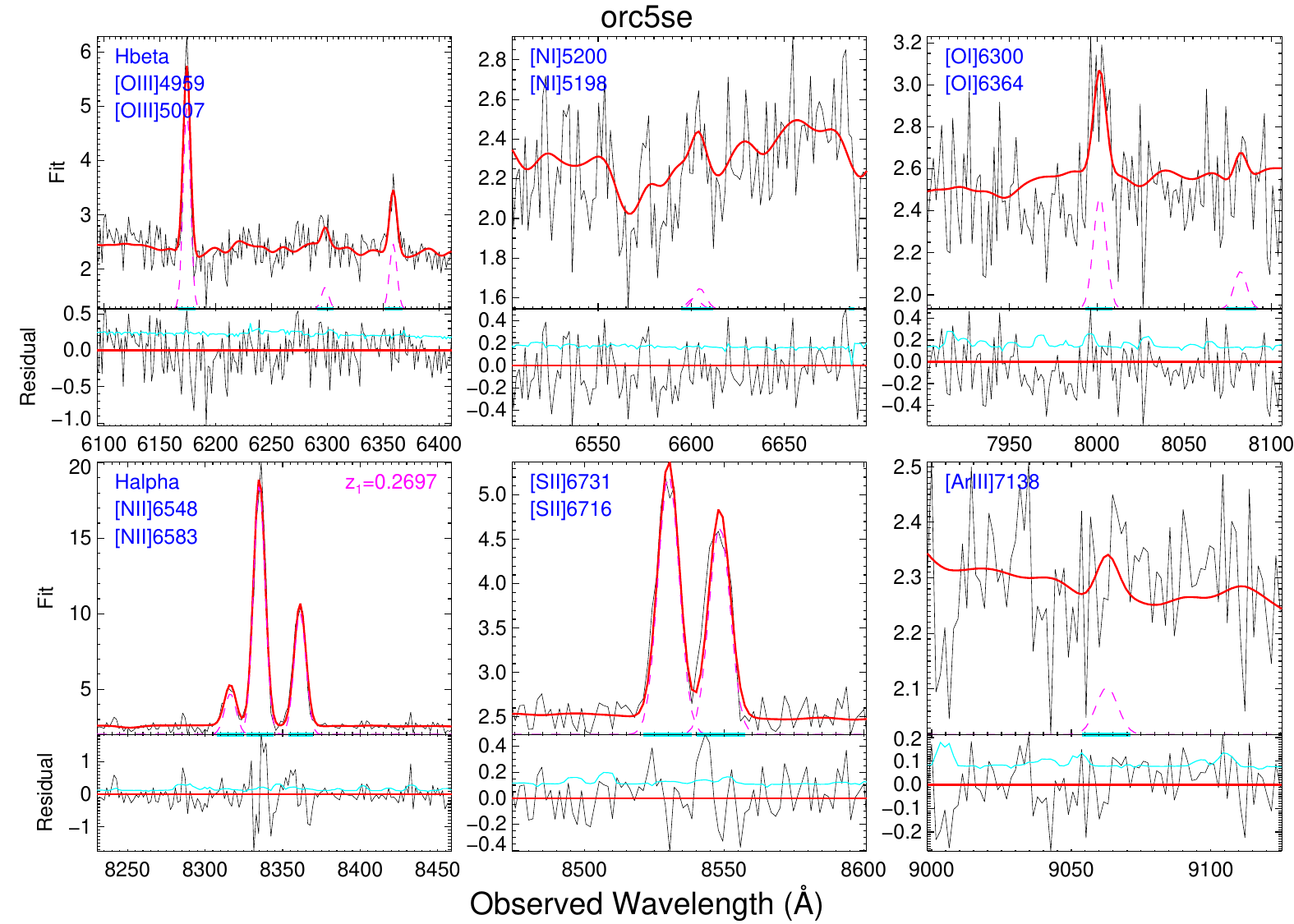}
    \caption{{\it Continued.}}
\end{figure*}
We fit the spectra of galaxies with identifiable redshifts using IFSFIT \citep{2014ascl.soft09005R,2017ApJ...850...40R}. IFSFIT first masks emission lines; fits a continuum model; and then fits the resulting continuum-subtracted emission-line spectrum. We use {\ttfamily ppxf} \citep{2012ascl.soft10002C,2017MNRAS.466..798C} to fit the stellar continua using the C3K theoretical stellar libraries at solar metallicity \citep{2016ApJ...823..102C, 2023MNRAS.521.4995B}. We allow for stellar attenuation in {\ttfamily ppxf}. We also include an additive Legendre polynomial of order 7 with positive-definite coefficients to account for small errors in flux calibration and/or slit losses. We estimate errors in the stellar fit parameters by refitting Monte Carlo realizations of the data. We fit emission lines with a single component and require a 2.5$\sigma$ detection in \ha\ and/or \otl. We estimate emission-line errors from the covariance matrix. Table~\ref{tab:spec-prop} lists the stellar and gas redshifts and velocity dispersions we measure from the fits, as well as the stellar color excess. We estimate an inferred black hole mass from $\sigma_*$ using the \citet{2013ApJ...764..184M} prescription and also list this in Table~\ref{tab:spec-prop}.

For visual inspection of the results of these fits, Figure~\ref{fig:spec-lines} shows close-up views of the regions around select emission lines.

%\movetabledown=in
%\LTcapwidth=\linewidth
%\begin{longrotatetable}
%\begin{deluxetable*}{lCCCCcCDCCrrr}
\begin{splitdeluxetable*}{lCCCCcDBCDCCCrrrC}
\tablecaption{Galaxy Properties Derived from Spectra\label{tab:spec-prop}}
  %\tabletypesize{\scriptsize}
  \tablewidth{0pt}
  \tablehead{\colhead{Label} & \colhead{$z_*$} & \colhead{$z_\mathrm{gas}$} & \colhead{$\sigma_*$} & \colhead{$\sigma_\mathrm{gas}$} & \colhead{Spec. Type} & \multicolumn2c{log\,L$_{[\mathrm{OIII}]}$} & \colhead{log\,L$_{\ha}$} & \multicolumn2c{SFR} & \colhead{\ebv$_*$} & \colhead{\ebv$_\mathrm{gas}$} & \colhead{D$_n$(4000)} & \colhead{young} & \colhead{int.} & \colhead{old} & \colhead{log\,M$_\mathrm{BH}$} \\
  \colhead{} & \colhead{} & \colhead{} & \colhead{\kms} & \colhead{\kms} & \colhead{} & \multicolumn2c{erg~s$^{-1}$} & \colhead{erg~s$^{-1}$} & \multicolumn2c{\smpy} & \colhead{mag} & \colhead{mag} & \colhead{} & \colhead{\%} & \colhead{\%} & \colhead{\%} & \colhead{\msun}}
  \colnumbers
  \decimals
  \startdata
   ORC1c & 0.54652\pm0.00005 & 0.54766\pm0.00010 & 249\pm13 &  \nodata &      LINER & 40.28^{+0.14}_{-0.21} & 40.54^{+0.17}_{-0.29} & <0.3^{+0.2} &  0.02\pm0.01 &      \nodata & 1.66\pm0.04 &   0 &  14 &  82 &  8.86\pm0.33  \\
  ORC23c & 0.25213\pm0.00008 & 0.25262\pm0.00002 & 196\pm22 & 121\pm6  &        AGN & 41.91^{+0.04}_{-0.05} & 41.68^{+0.03}_{-0.04} & <3.7^{+0.0} &  0.11\pm0.02 &      \nodata & 1.04\pm0.47 &   0 &   0 &  99 &  8.28\pm0.41  \\
   ORC4c & 0.45089\pm0.00004 & 0.45115\pm0.00002 & 227\pm11 & 172\pm2  &      LINER & 41.44^{+0.04}_{-0.04} & 41.81^{+0.03}_{-0.03} & <5.1^{+0.0} &  0.00\pm0.01 &  0.15\pm0.05 & 1.30\pm0.05 &   7 &   0 &  79 &  8.64\pm0.33  \\
  ORC4se & 0.44914\pm0.00006 &           \nodata & 166\pm20 &  \nodata &    \nodata &               \nodata &               \nodata &     \nodata &  0.11\pm0.04 &      \nodata & 2.34\pm0.03 &   0 &   0 & 100 &  7.88\pm0.42  \\
   ORC5c & 0.26887\pm0.00004 & 0.26956\pm0.00013 & 233\pm11 & 142\pm31 &  HII/LINER &        <39.40^{+0.72} & 40.59^{+0.10}_{-0.13} & <0.3^{+0.1} &  0.00\pm0.01 &      \nodata & 2.29\pm0.29 &   0 &   0 & 100 &  8.70\pm0.32  \\
  ORC5se & 0.26961\pm0.00008 & 0.26969\pm0.00000 & 263\pm33 & 100\pm1  &        HII & 41.85^{+0.06}_{-0.07} & 42.85^{+0.03}_{-0.03} &  56.5\pm3.7 &  0.19\pm0.02 &  0.55\pm0.05 & 1.44\pm0.25 &  43 &   0 &  57 &  9.00\pm0.43  \\
  \enddata
  \tablecomments{Column 1: Label from Figure~\ref{fig:img-slit}. Columns 2--5: Stellar and ionized gas redshifts and velocity dispersions. Errors shown are 1$\sigma$. Redshifts are heliocentric and velocity dispersions intrinsic (corrected for instrumental effects). Column 6: Adopted spectral type. Columns 7--8: Extinction-corrected \othl\ and \ha\ luminosities. Where nebular extinction is not available, we correct using $\ebv_\mathrm{gas} = \ebv_*/0.44$ \citep{2000ApJ...533..682C}. Column 9: Star formation rate from \ha\ \citep{2006ApJ...642..775M}. Set to upper limit if not an HII spectral type. Column 10: Stellar broadband colour excess derived from \texttt{ppxf} fits that use the \citet{2000ApJ...533..682C} attenuation curve with $R_V = 4.05$. We estimate the 1$\sigma$ error with Monte Carlo simulations. Column 11: Nebular broadband colour excess and 1$\sigma$ error, derived from \ha/\hb\ decrement assuming Case B conditions, the \citet{1989ApJ...345..245C} extinction curve, and $R_V = 4.05$. Column 12: The Balmer break index as defined in \citet{1999ApJ...527...54B}. Columns 13--15: Percentage contribution to the total starlight model at rest-frame 5000~\AA\ in bins of young ($<$100~Myr), intermediate-age ($100~\mathrm{Myr}-1~\mathrm{Gyr}$), and old ($>$1~Gyr) populations. The sum may be less than 100\%\ if there is an additive polynomial contribution. Column 16: Black hole mass estimated from $\sigma_*$ \citep{2013ApJ...764..184M}. Error includes measurement error in $\sigma_*$ and the scatter in the $M_\mathrm{BH}-\sigma$ relation.}  
\end{splitdeluxetable*}
\begin{figure*}[t]
    \centering
    \includegraphics[width=\textwidth]{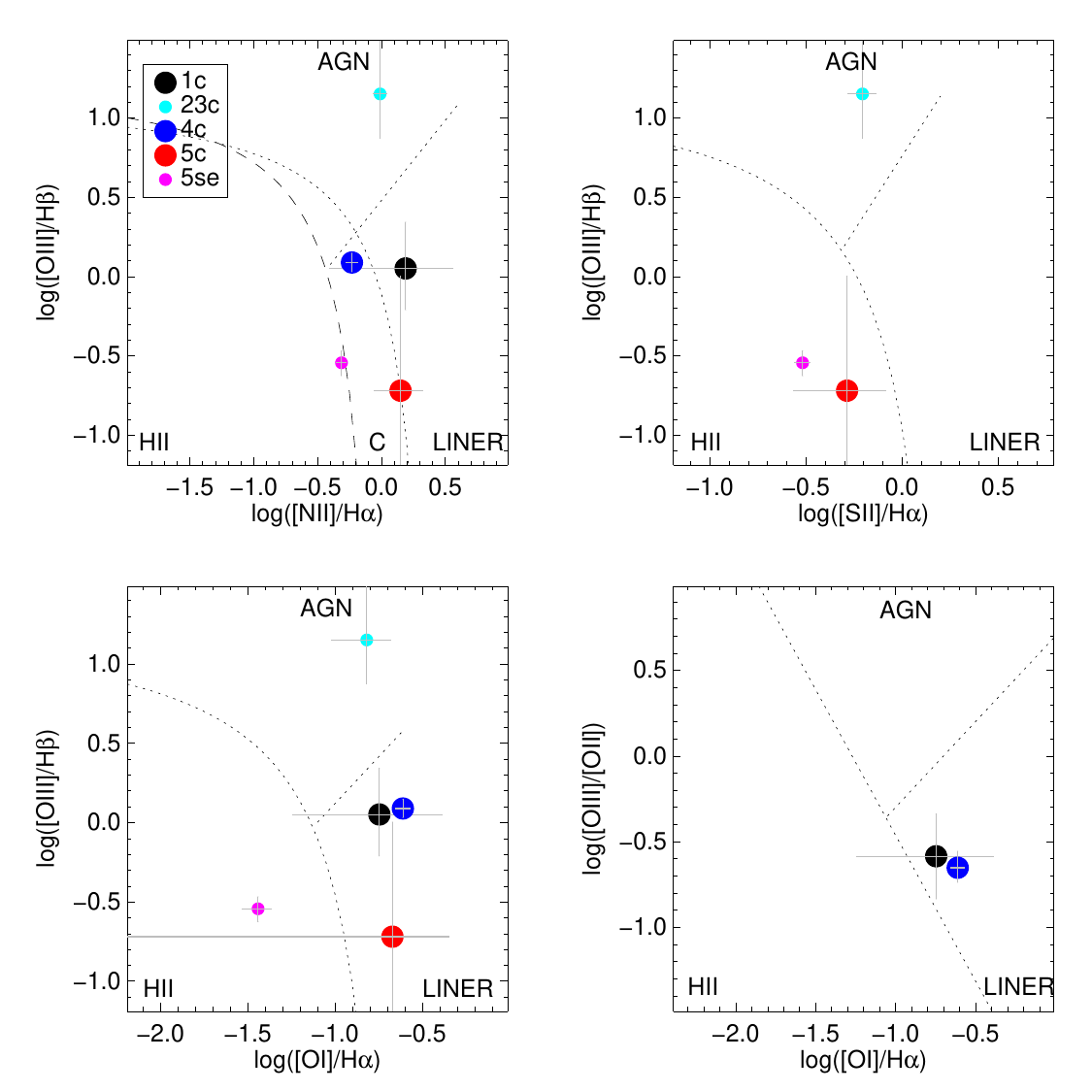}
    \caption{Standard excitation diagrams \citep{1981PASP...93....5B,1987ApJS...63..295V,2006MNRAS.372..961K} for the five identified galaxies with emission lines. The dashed and dotted lines divide nuclear classification regions of star-forming (\ion{H}{2}), AGN, Low-Ionization Nuclear Emission Line Region (LINER), and composite (C) using canonical dividing lines \citep{2003MNRAS.346.1055K,2006MNRAS.372..961K,2010MNRAS.403.1036C}. The adopted spectral type of each source is listed in Table~\ref{tab:spec-prop}.}
    \label{fig:vo}
\end{figure*}
To parameterize the stellar ages, we quantify the percentage contribution to the total starlight model at a rest-frame wavelength of 5000~\AA\ in bins of young ($<$100~Myr), intermediate-age ($100~\mathrm{Myr}-1~\mathrm{Gyr}$), and old ($>$1~Gyr) populations from the stellar libraries used to model the continuum. These percentages are listed in Table~\ref{tab:spec-prop}. (The sum may be less than 100\%\ if there is an additive polynomial contribution.) We also calculate the D$n$(4000) index to parameterize the Balmer break \citep{1999ApJ...527...54B}, which monotonically increases with the age after an instantaneous burst \citep{2003MNRAS.341...33K}.

For spectra in which H$\beta$ is detected at 2$\sigma$, we estimate the color excess \ebv\ from the Balmer decrement, assuming case B, the \citet{1989ApJ...345..245C} extinction law, and $R_V=4.05$. We then correct line fluxes for extinction and compute the line ratios needed to construct standard excitation diagrams \citep{1981PASP...93....5B,1987ApJS...63..295V,2006MNRAS.372..961K}. For galaxies without measured nebular extinction, we correct fluxes using the stellar color excess divided by 0.44 \citep{2000ApJ...533..682C}. We list the estimated nebular color excess in Table~\ref{tab:spec-prop}, as well as the intrinsic \oth\ and \ha\ luminosities. Figure~\ref{fig:vo} shows the excitation diagrams with lines separating emission-line classes; we list the adopted spectral types in Table~\ref{tab:spec-prop}. For the \ion{H}{2} galaxy we compute SFR from \ha\ \citep{2006ApJ...642..775M}. For other galaxies, we compute a SFR upper limit, as at least some of the \ha\ emission arises from processes other than photoionization from young stars.

The excitation diagrams show that the spectral types of ORC1c and ORC4c are most consistent with being low-ionization nuclear emission line regions (LINERs). ORC4c lies in the composite region of the  \nt/\ha\ versus \oth/\hb\ diagram (hereafter n2ha), but is squarely in the LINER region of the \oo/\ha\ versus \oth/\hb\ (o1ha) and \oo/\ha\ versus \oth/\ot\ (o3o2) diagrams. \oo\ is more diagnostic of LINER regions because it traces the extended partially-ionized zone of nebulae, and has previously been used to define LINER sub-classes \citep{2010MNRAS.402.2187S}. The low-ionization \noll\ line has also been used to categorize LINERs in nearby ellipticals \citep{2010MNRAS.402.2187S}. In ORC4c, $\no/\hb=-0.59\pm0.17$, which in combination with \oth/\hb\ puts it in the same LINER class as other $z=0$ massive ellipticals \citep{2010MNRAS.402.2187S}.

ORC23c and ORC5se are unambiguously AGN and star-forming nuclei, respectively. We list ORC5c as an \ion{H}{2}/LINER, as its lines are intrinsically weak and the line ratios are uncertain, particularly \oth/\hb\ and \oo/\ha. Its placement in the n2ha, s2ha, and o1ha diagrams are consistent with LINER/composite, \ion{H}{2}, and \ion{H}{2}/LINER within the 1$\sigma$ errors, respectively.

\subsection{Spectral Energy Distributions} \label{sec:results-sed}

As we did in \citet{2024Natur.625..459C} for ORC4c, we use the Tractor photometry to perform Prospector \citep{2021ApJS..254...22J} fits to the rest-frame optical-to-near-infrared spectral energy distribution of each source with a known redshift. We fit to the observed-frame $griz$ and W1/W2 photometry. Prospector infers the properties of the galaxy's stars through gridless Bayesian parameter estimation and Markov chain Monte Carlo (MCMC) posterior sampling with \texttt{emcee} \citep{2013PASP..125..306F}. The starlight models are generated with Flexible Stellar Population Synthesis (FSPS; \citealt{2009ApJ...699..486C,2010ascl.soft10043C}). The goal of these fits is not to reconstruct an accurate star formation history for each source, but rather to estimate the stellar mass, to get some constraints on current and recent star formation, and to put further constraints on any current AGN activity.

We assume a Kroupa initial mass function \citep{2001MNRAS.322..231K} and a power law attenuation $A_\lambda/A_V \sim \lambda^{-0.7}$ that matches the Milky Way's UV-to-optical slope \citep{2020ARA&A..58..529S}. To parameterize the star formation history, we use a delayed $\tau$-model such that SFR declines as $t_\mathrm{age}\,e^{-t_\mathrm{age}/\tau}$, where $t_\mathrm{age}$ is the galaxy's age. We set a log-uniform prior on $t_\mathrm{age}$ such that the galaxy is not younger than 100 Myr when observed and doesn't form before the Universe ($0.1~ \mathrm{Gyr} < t_\mathrm{age} < t_\mathrm{Universe}(z)$, where $t_\mathrm{Universe}(z)$ is the age of the Universe at the galaxy's redshift $z$). We allow the bolometric fraction of the light due to Type I AGN to be a free parameter. The stellar mass $M_*$ is a free parameter, with a log-normal prior with a mode of ln$(M_*/\msun) = 25$ and a dispersion of 15. Finally, we update the stellar metallicity for each random draw of M$_*$ to be within $\pm$0.1 dex of the mass-metallicity relation \citep{2016MNRAS.456.2140M}.

For ORC~5se, which is clearly a star-forming galaxy rather than an AGN (Section~\ref{sec:discuss-sources}), we set the AGN fraction to zero and allow a recent burst.

We list the best-fit parameters and 1$\sigma$ errors in Table~\ref{tab:sed-prop}. We also compute and list the current star formation rate and star formation rate 1~Gyr in the past using
\begin{eqnarray}
    \mathrm{SFR} &\,=\,& \mathrm{SFR_{max}}\frac{t_\mathrm{age}}{\tau}(e^{1-t_\mathrm{age}/\tau}) \\
  \mathrm{SFR_{max}} &\,=\,& M_*(e^{-1}\times10^{-9}~\smpy)\;\times \nonumber \\
  ~ & ~ & \tau^{-1}[1-(1+\frac{t_\mathrm{age}}{\tau})e^{-t_\mathrm{age}/\tau}]^{-1}
\end{eqnarray}
where $t_\mathrm{age}$ and $\tau$ are in Gyr and M$_*$ is in \msun.

This differs from our methodology in \citet{2024Natur.625..459C}, in which we added both an AGN and burst to the $\tau$-model for ORC4c. In that fit, we incorporated additional $u$-band data to allow for a more detailed star formation history. However, the results are consistent--the inferred masses are the same, and the \citet{2024Natur.625..459C} fit finds a burst about 1~Gyr in the past--like the elevated SFR we find 1~Gyr in the galaxy's past.
% SFH: delayed tau model + AGN
\begin{deluxetable*}{lCDCcDC}
  \tablecaption{Galaxy Properties Derived from SED Fits\label{tab:sed-prop}}
  %\tablewidth{0pt}
  \tablehead{\colhead{Label} & \colhead{log($M_*$/\msun)} & \multicolumn2c{$t_\mathrm{age}$} & \colhead{$\tau$} & \colhead{SFR$_\mathrm{now}$} & \multicolumn2c{SFR$_\mathrm{1~Gyr}$} & \colhead{AGN fraction} \\
    & & \multicolumn2c{Gyr} & \colhead{Gyr} & \multicolumn2c{\smpy} & \colhead{\smpy} & }
  \colnumbers
  \decimals
  \startdata
   ORC1c & 11.7\pm0.3 & 1.52^{+0.05}_{-0.03} & 0.02^{+0.02}_{-0.01} & 0.0 & 0.0 & 0.01^{+0.01}_{-0.01} \\
   ORC23c & 11.2\pm0.3 & 10.3^{+0.2}_{-0.4} & 1.40^{+0.03}_{-0.06} & 0.5 & 1.0 & 0.18^{+0.03}_{-0.03} \\
   ORC4c & 11.4\pm0.3 & 2.3^{+0.2}_{-0.2} & 0.26^{+0.04}_{-0.04} & 1.2 & 33 & 0.00^{+0.01}_{-0.00} \\
   ORC4se & 10.6\pm0.3 & 3.4^{+0.8}_{-0.5} & 0.23^{+0.17}_{-0.15} & 0.0 & 0.0 & 0.06^{+0.10}_{-0.05} \\
   ORC5c & 11.2\pm0.3 & 2.98^{+0.06}_{-0.07} & 0.19^{+0.02}_{-0.02} & 0.0 & 0.3 & 0.00^{+0.00}_{-0.00} \\
   ORC5se & 11.0\pm0.3 & 10.17^{+0.11}_{-0.27} & 1.98^{+0.06}_{-0.07} & \nodata\tablenotemark{a} & \nodata\tablenotemark{a} & 0.0\tablenotemark{a} \\
  \enddata
  \tablecomments{Column 1: Label. Column 2: Best-fit stellar mass from Prospector fits; we assign a conservative factor-of-two error. Columns 3--4: Best-fit parameters for a delayed-$\tau$ star formation history $t_\mathrm{age}\,e^{-t_\mathrm{age}/\tau}$. Columns 5--6: Current SFR and SFR 1~Gyr in the past from best-fit star formation history. Column 7: Best-fit AGN fraction.}
    \tablenotetext{a}{For this galaxy we do not include an AGN and instead add a burst to the model. The best-fit burst parameters are for 33$^{+0.03}_{-0.05}$\%\ of the galaxy mass to have formed at $0.84\pm0.01$ of the galaxy's current age, which is roughly 1.6~Gyr in the past.}
\end{deluxetable*}

\section{Discussion} \label{sec:discuss}

\subsection{Characterizing the galaxies of ORCs 1--5} \label{sec:discuss-sources}

We are now in a position to characterize the properties of the central galaxies of ORCs, as well as other, possibly related, galaxies in or near the ORC footprints. This provides stronger constraints on the possible origin of these phenomena, as well as the intrinsic sizes of the circles.

The three isolated ORCs in our study (ORC1, ORC4, and ORC5) each have a central source well outside the local Universe ($z = 0.547$, 0.451, and 0.270). These measurements confirm or refine initial estimates from photometry \citep{2021PASA...38....3N,2021MNRAS.505L..11K,2022MNRAS.513.1300N}. For ORC4c, the initial photometric redshift estimate of $z=0.385$ was significantly smaller than the spectroscopic redshift \citep{2024Natur.625..459C}.

Besides their distances, the systems ORC1c, ORC4c, and ORC5c are remarkably similar in other ways. They are red elliptical galaxies with no internal stellar attenuation (Tables~\ref{tab:coord-shape} and \ref{tab:spec-prop}). Following \citet{2022MNRAS.513.1300N}, we verify that their $g-r$, $r-i$, $W1-W2$, and $W2-W3$ colors\footnote{We compute WISE colors in the Vega system for literature comparison using \citet{2012wise.rept....1C}.} (Table~\ref{tab:phot}) are consistent with early type galaxies, with no evidence for AGN activity \citep{2019ApJS..245...25J,2020ApJ...903...91Y}. Their light-weighted stellar populations are dominated by stars with ages $>$1~Gyr (80-100\%\ of the light at rest-frame 5000~\AA; Table~\ref{tab:spec-prop}). They have velocity dispersions $225-250$~\kms\ (Table~\ref{tab:spec-prop}), roughly consistent with the stellar masses from SED fitting of M$_* =2-5\times10^{11}$~\msun\ (Table~\ref{tab:sed-prop}). They each have some nebular emission, and spectral types consistent with LINER (Table~\ref{tab:spec-prop}). Based on the Prospector fits, there is no evidence for a luminous AGN in the optical and near-infrared (Table~\ref{tab:sed-prop}). 

ORC1c and ORC5c have H$\alpha$ luminosities (log$[L_{\ha}/\lsun] = 40.5-40.6$; Table~\ref{tab:spec-prop}) that are about 10$\times$ higher than $z=0$ ellipticals of comparable mass \citep{2006MNRAS.366.1151S,2006MNRAS.366.1126C}. The excess could arise due to shocks, as we discuss below. However, even if they were powered by stellar photoionization, the implied SFR is negligible (Table~\ref{tab:spec-prop}). The WISE W3 non-detections give 1$\sigma$ upper limits of 15 and 4~\smpy, respectively \citep{2017ApJ...850...68C}, that are less stringent. Their SEDs are consistent with little star formation over the last Gyr (Table~\ref{tab:sed-prop}), as are their Balmer break D$_n$(4000) indices (Table~\ref{tab:spec-prop}; \citealt{2003MNRAS.341...33K}).

ORC4c is unique among these three central galaxies in that its H$\alpha$ emission is about 10$\times$ brighter than that of ORC1c and ORC5c (Table~\ref{tab:spec-prop}). In the \otl\ line, it is particularly luminous ($W_\mathrm{eq} = 50$~\AA), extended over 40~kpc, and turbulent ($\sigma_\mathrm{gas}=180$~\kms; \citealt{2024Natur.625..459C}). With our longslit data, we recover about 80\%\ of the flux observed with KCWI and a similar velocity dispersion (Table~\ref{tab:spec-prop}; \citealt{2024Natur.625..459C}). In comparison, the emission lines in ORC1c are unresolved, while those in ORC5c have $\sigma_\mathrm{gas}=142\pm31$~\kms\ (Table~\ref{tab:spec-prop}). The star formation rate in ORC4c is low, with $\mathrm{SFR}<5-7$~\smpy\ from H$\alpha$ (Table~\ref{tab:spec-prop}) and WISE W3 \citep{2017ApJ...850...68C}. However, the SFR may have been significantly elevated sometime in the last Gyr. The spectrum indicates some contribution from young stars (Table~\ref{tab:spec-prop}), and the parameterized star-formation history from SED modeling shows a decrease in the SFR from 30 to 1~\smpy\ over the past Gyr (Table~\ref{tab:sed-prop}). Using additional $u$-band data, the more detailed star-formation history of ORC1c presented in \citet{2024Natur.625..459C} is consistent with an intense starburst about 1~Gyr ago, as is its lower D$_n$(4000) value  (Table~\ref{tab:spec-prop}; \citealt{2003MNRAS.341...33K}).

\begin{figure*}
    \centering
    \includegraphics[width=\textwidth]{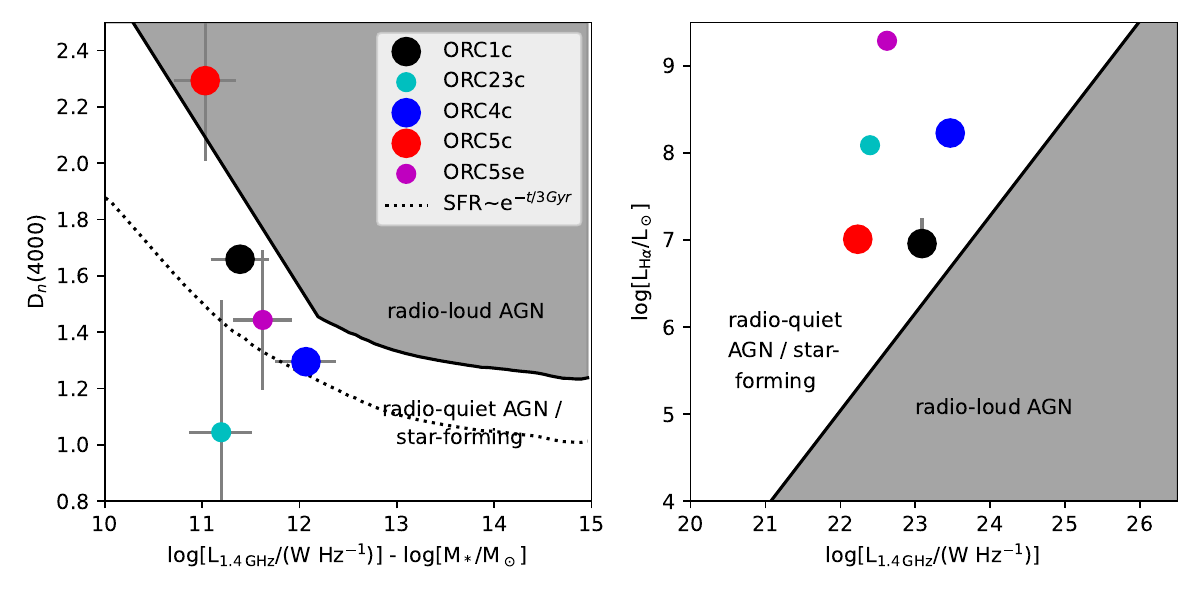}
    \caption{Two parts of the radio-based galaxy classification scheme of \citet{2012MNRAS.421.1569B}; the third uses the n2ha excitation diagram (Figure~\ref{fig:vo}). The dashed line is the evolutionary track of a galaxy with an exponentially-declining star formation rate with e-folding time 3~Gyr \citep{2003MNRAS.344.1000B}. The solid lines divide radio-loud AGN from radio-quiet AGN and star-forming galaxies. We digitized the dividing line in the D$_n$(4000) vs. $L_{1.4}-M_*$ diagram from \citet{2012MNRAS.421.1569B} using WebPlotDigitzer \citep{Rohatgi2022}.}
    \label{fig:best}
\end{figure*}

The radio emission in these galaxies is in excess of that expected from star formation. We calculate the rest-frame 1.4~GHz flux density using published measurements and spectral indices (Table~\ref{tab:rad-prop}). \citet{2022MNRAS.513.1300N} and \citet{2021MNRAS.505L..11K} argue for radio AGN in ORC1c, ORC4c, and ORC5c based on the differences between the star formation rate estimated from the radio and other tracers. Our star formation upper limits based on H$\alpha$ are about 10$\times$ lower than those of \citet{2022MNRAS.513.1300N}, and thus strengthen this conclusion. They imply that the 1.4~GHz radio emission is brighter than expected from star formation in these sources by large factors. Because the H$\alpha$ in these sources are upper limits, the lower limits on the 1.4~GHz excess ($L_{1.4}/L_{1.4,\,\mathrm{SFR}}$) are 150--430 for ORC 1c and 20--50 for ORCs 4c and 5c. Radio sources like ORC1c, ORC4c, and ORC5c that do not have strong high-ionization optical lines or an excess of hot dust (based on WISE colors), but that do have a significant radio-to-infrared excess, may comprise a significant number of true AGNs \citep{2022ApJ...939...26Y}.
\begin{deluxetable*}{lcDDcCCCD}
  \tablecaption{Radio Properties\label{tab:rad-prop}}
  \tablewidth{0pt}
  \tabletypesize{\scriptsize}
  \tablehead{\colhead{Label} & \colhead{$\nu_\mathrm{obs}$} & \multicolumn2c{$S_{\nu,\,\mathrm{obs}}$} & \multicolumn2c{$\alpha$} & \colhead{$\nu_{1.4}$} & \colhead{$S_{\nu,\,1.4}$} & \colhead{$L_{1.4}$} & \colhead{$L_{1.4}/L_{1.4,\,\mathrm{SFR}}$} & \multicolumn2c{log($L/L_\mathrm{Edd}$)} \\
    & \colhead{GHz} & \multicolumn2c{mJy} & \multicolumn2c{} & \colhead{GHz} & \colhead{mJy} & \colhead{10$^{22}$ W Hz$^{-1}$} & \colhead{} & \multicolumn2c{}}
  \colnumbers
  \decimals
  \startdata
ORC1c  & 1.284 & 0.091\pm0.004 & -1.40\pm 0.05 & 0.905 & 0.15\pm0.01 & 12.2\pm 0.7 & >292\pm142 & -3.74\pm0.37 \\
ORC23c & 0.944 &  0.19\pm 0.05 & -1.23\pm 0.36 & 1.118 & 0.15\pm0.04 &  2.6\pm 0.7 & >  4\pm  1 & -2.30\pm0.46 \\
ORC4c  & 0.325 &  1.43\pm 0.13 & -0.92\pm 0.18 & 0.965 & 0.53\pm0.21 &   29\pm  12 & > 37\pm 15 & -2.93\pm0.34\tablenotemark{a} \\
ORC5c  & 0.944 &  0.10\pm 0.03 &  -0.8\pm  0.2 & 1.103 & 0.09\pm0.03 &  1.7\pm 0.5 & > 35\pm 14 & <-4.20^{+0.78} \\
ORC5se & 0.944 &  0.25\pm 0.03 &  -0.8\pm  0.2 & 1.103 & 0.22\pm0.03 &  4.2\pm 0.6 & 0.49\pm0.07 & \nodata \\
\enddata
  \tablecomments{Column 1: Label. Column 2--4: Observed frequency, flux density, and spectral index \citep{2021PASA...38....3N,2021MNRAS.505L..11K,2022MNRAS.513.1300N}, where $S_\nu \sim \nu^\alpha$. Where not explicitly listed in the literature, flux errors are assumed to be twice the r.m.s. sensitivity per beam. Column 5--6: Observed frequency corresponding to 1.4~GHz in the galaxy's rest frame and flux density at this frequency extrapolated from $S_{\nu,\,1.4} = S_{\nu,\,\mathrm{obs}}(\nu_\mathrm{obs}/\nu_{1.4})^\alpha$. Column 7: Luminosity density at 1.4~GHz in the galaxy's rest frame, in units of 10$^{22}$ W~Hz$^{-1}$: $L_{1.4} = 4\pi d_L^2\,S_{\nu,\,1.4}/(1+z)$, where $d_L$ is the luminosity distance. Column 8: Excess 1.4~Ghz emission above that expected from star formation. The SFR is from \ha\ (Table~\ref{tab:spec-prop}), which provides only upper limits for most sources. We convert SFR to $L_{1.4,\,\mathrm{SFR}}$ using the calibrataion from \citet{2021ApJ...914..126M}, assuming a Salpeter IMF to match the \ha\ SFR calibration of \citet{2006ApJ...642..775M}. The errors are propagated from the fluxes and do not include systematics in the calibration. Column 9: Eddington ratio, where the observed luminosity combines radiation and jet power. We use an \othl\ bolometric correction of 142 \citep{2009AA...504...73L} for radiation and equation (2) of \citet{2014ARAA..52..589H} for the radio jet power.}
\tablenotetext{a}{This is an upper limit if shocks contribute significantly to $L_\mathrm{[OIII]}$ and/or $L_{1.4}$.}
\end{deluxetable*}
LINER-like emission can arise from multiple physical processes. We argue below that the emission in ORC4c, and perhaps also 1c and 5c, is shock-dominated (Section~\ref{sec:discuss-origin}). However, the LINER classifications of ORCs 1c, 4c, and 5c are also consistent with an AGN in these sources. We combine our optical data with the radio results using the AGN classification scheme of \citet{2012MNRAS.421.1569B}. In Figure~\ref{fig:best}, we show where these galaxies fall in two parts of this scheme; the third is based on the n2ha diagram (Figure~\ref{fig:vo}). The scheme places these galaxies in the radio-quiet AGN class. (Formally, ORC~5c is in the radio-loud AGN class, but its D$_n$(4000) has a large error bar that overlaps the radio-quiet class; its low radio luminosity is more consistent with the radio-quiet class.) Following \citet{2012MNRAS.421.1569B}, we compute the Eddington ratio in these sources by combining the AGN power in radiation and mechanical jet power. We apply a bolometric correction of 142 to $L_\mathrm{[OIII]}$ \citep{2009AA...504...73L} and use the calibration of \citet{2014ARAA..52..589H} to compute the radio jet power. The result is a low Eddington ratio: log\,$(L/L_\mathrm{Edd}) = -3$ to $-4$. These values are consistent with radiatively-inefficient accretion, as observed in lower-luminosity AGN \citep{2009ApJ...699..626H}. The Eddington ratio  of these systems is even lower if the line (and/or radio) emission is dominated by shocks, an effect which is likely strongest in ORC~4c.
Our spectra uncover two other galaxies within the radio footprints of ORCs 4 and 5 that lie at the same redshift as the central source. ORC4se is another old, red elliptical (Tables~\ref{tab:coord-shape}, \ref{tab:spec-prop}, and \ref{tab:sed-prop}) with no nebular emission or star formation detected (Table~\ref{tab:spec-prop}). It lies at a projected distance of 35~kpc from ORC4c (Table~\ref{tab:coord-shape}), and is about 6$\times$ less massive ($\sigma_* = 166\pm18$~\kms, log[$M_*/\msun]=10.6$; Tables~\ref{tab:spec-prop} and \ref{tab:sed-prop}). ORC5se lies at the edge of ORC5 radio ring, at a projected distance of 151~kpc from ORC5c (Table~\ref{tab:coord-shape}). It is a spheroidal system (Table~\ref{tab:coord-shape}) classified as a star-forming galaxy (Figure~\ref{fig:vo}). Its spectrum has a significant contribution from a young, modestly obscured stellar population (Table~\ref{tab:spec-prop}). Like ORC5c, it is massive ($\sigma_* = 263\pm33$~\kms\ and log[$M_*/\msun]=11.0$; Tables~\ref{tab:spec-prop} and ~\ref{tab:sed-prop}). Its star formation rate is 57~\smpy\ based on H$\alpha$ (Table~\ref{tab:spec-prop}), $22\pm16$~\smpy\ based on W3 \citep{2017ApJ...850...68C}, and $27\pm3$ from radio emission (Table~\ref{tab:rad-prop}). The SED fit points to a burst in the last 2~Gyr (Table~\ref{tab:sed-prop}), consistent with the D$_n$(4000) value  (Table~\ref{tab:spec-prop}; \citealt{2003MNRAS.341...33K}), but is not sensitive to the ongoing star formation.

ORCs 2 and 3 have no central galaxies. We observed the system directly between them, which is a disk galaxy at $z=0.252$ (Table~\ref{tab:spec-prop}), somewhat closer than the previous photometric estimate ($z = 0.286$; \citealt{2021PASA...38....3N}). We discuss below whether ORC23c is related to ORCs 2 and 3. However, if we do associate ORC23c with ORCs 2 and 3 and assume they all lie at the same redshift, then ORC23c is 168~kpc in projection from the center of ORC2 and 274~kpc from the center of ORC3 (Table~\ref{tab:coord-shape} and \citealt{2021PASA...38....3N}). This system also hosts an old stellar population (Table~\ref{tab:spec-prop}), but in this case with a Type 2 AGN (Figure~\ref{fig:vo}). Its bolometric luminosity (based on $L_\mathrm{[OIII]}$; Table~\ref{tab:spec-prop}) and Eddington ratio (Table~\ref{tab:rad-prop}) are the highest in the sample. It is firmly in the radio-quiet AGN category (Figure~\ref{fig:best}; \citep{2012MNRAS.421.1569B}) along with ORC1c, 4c, and 5c, but with a higher radiative vs. jet output.

\subsection{Origins of the ORC phenomenon} \label{sec:discuss-origin}

The confirmed redshifts of the central galaxies of ORCs 1, 4, and 5 rule out a Galactic origin. The intrinsic diameters of the ORCs inferred from these redshifts are thus in the range 300-500~kpc, as previously inferred based on photometric redshifts \citep{2022MNRAS.513.1300N}. They instead point to one of the various mechanisms that have been proposed that involve galaxy-wide processes. These include a powerful wind shock propagating into the CGM \citep{2022MNRAS.513.1300N,2024Natur.625..459C}, shocks driven by the gravitational motions in a galaxy merger \citep{2023ApJ...945...74D}, or a virial shock \citep{2024MNRAS.528.3854Y}. 

In \citet{2024Natur.625..459C}, we argue for a galactic wind model for ORC4c based on the morphology, kinematics, and luminosity of the extended line emission. We also show that a 2D hydrodynamic model of a blast wave driven for 200~Myr by a 200~\smpy\ starburst, or alternatively a luminous AGN, can reproduce the optical and radio data. In this scenario, the large-scale radio emission is produced by the forward shock and observed 800~Myr after the wind begin. The ionized gas was reheated by the reverse wind shock and by further shocks as it fell back on to the central galaxy.

It is likely that in ORC4c, the line emission---and possibly also the central radio emission---have a dominant contribution from extended shocks. Its line ratios and line widths are consistent with those seen, for instance, in the extended regions of the giant relic wind around the starburst galaxy Makani \citep{2023ApJ...947...33R}. The rest-frame equivalent width of the \otl\ emission in this source is 50~\AA\ \citet{2024Natur.625..459C}, which is 10$\times$ higher than in other massive ellipticals \citep{2017ApJ...837...40P}. If shocks dominate $L_{1.4}$ and/or $L_\mathrm{[OIII]}$, the estimated Eddington ratio is an upper limit (Table~\ref{tab:rad-prop}).

Though they overlap with ORC4c in excitation diagrams (Figure~\ref{fig:vo}), the origin of the LINER emission in ORC1c and 5c is murkier due to their lower line luminosities and widths. One possibility is shocks in a relic wind, as we hypothesize for ORC4c. Others include photoionization and shock emission from a low-luminosity AGN \citep[e.g.,][]{2018ApJ...864...90M} or photoionization by evolved stars \citep[e.g.,][]{2010MNRAS.402.2187S}. The \ha\ luminosities of ORC1c and ORC5c are $\sim$10$\times$ smaller than that of ORC4c (Table~\ref{tab:spec-prop}), though still 10$\times$ higher than comparably massive, $z=0$ ellipticals \citep{2006MNRAS.366.1151S}. Their line equivalent widths are also lower than in ORC4c; we measure $W_\mathrm{eq}^\mathrm{rest} = 9~\AA$ for \ot\ in ORC1c, which is consistent with other massive galaxies \citep{2017ApJ...837...40P}. The central radio luminosity of ORC4c is higher than in ORC1c and ORC5c by factors of a few to $\sim$10 (Table~\ref{tab:rad-prop}), which could point to a larger shock contribution. Finally, the \oth\ equivalent widths of ORC1c and ORC4c differ by a factor of 5--10, consistent with a higher contribution of shocks to $L_\mathrm{[OIII]}$, as well, in ORC4c compared to ORC1c.

Despite these differences in line luminosity, the other similarities of the central galaxies of the three ORCs may point to a common origin. If it is in a starburst or AGN blast wave, then the current radio emission must have moved at modest speeds to the currently observed radii of 150--250~kpc. If it is due to a starburst, the burst appears to be in the distant past ($\ga$1~Gyr ago). The ballistic flow speed over 1~Gyr is 150--250~\kms\ for 150--250~kpc. This is remarkably similar to the inferred flow speeds for the clearly wind-powered nebula in Makani \citep{2019Natur.574..643R}. If the radio bubbles were instead powered by an AGN event, any evidence would have disappeared more quickly, and the constraints on timescale are weaker. That said, an energetic AGN event would probably also coincide with significant star formation \citep{2006ApJ...649...79S,2019MNRAS.484.4360A}, which we conclude is well in the past. In the blastwave scenario, we speculate that in ORC1c and ORC5c, the lower luminosity of the ionized gas results from reduced shock emission due to an event occurring further in the past.

Aside from the blastwave scenario, the current data do not provide further constraints on other extragalactic models, such as the galaxy merger \citep{2023ApJ...945...74D} or virial shock \citep{2024MNRAS.528.3854Y} models. However, the morphologies, stellar masses, stellar populations, and ionized gas properties point to massive, red elliptical central galaxies in ORCs 1, 4, and 5. These models feature such galaxies, either as a merger remnant or host of a massive halo.

The absence of a central galaxy points to a different origin for ORCs 2/3, as previously argued \citep{2021PASA...38....3N}. A scenario where ORCs 2/3 are the lobes of a radio galaxy (ORC23c) would be difficult to support if ORC23c were star-forming \citep{2021PASA...38....3N}. However, we find that ORC23c is actually a Type 2 AGN in a massive, red disk. We suggest that the double radio lobe scenario is therefore still possible in this case. If so, then the projected end-to-end scale of this radio galaxy is 550~kpc. 

Other difficulties with this scenario remain, in that the circularity of such lobes is not common, though such systems do exist \citep{2008MNRAS.390..595M}. For instance, as pointed out by \citet{2021PASA...38....3N}, the lobes of Fornax~A are also circular. The resemblence of Fornax A to ORCs 2/3 increases at lower resolution \citep{1983A&A...127..361E}, though the end-to-end scale of Fornax A's lobes is smaller at 200~kpc (using a distance of 20.0~Mpc from surface brightness fluctuations; \citealt{2009ApJ...694..556B}).

\section{Conclusion} \label{sec:conclude}

The Odd Radio Circles are an intriguing phenomena discovered in large-area radio surveys. Among origin theories, the possibility that they are relic remnants of energetic events, potentially representing the largest observed galactic winds, is compelling. Here we present optical spectroscopy of all five initial ORC discoveries, covering the three ORCs with central galaxies (1c, 4c, and 5c) and a galaxy located directly between ORCs 2 and 3 (ORC23c). We also observe two galaxies within the ORC 4 and 5 footprints (ORC4se and ORC5se), both of which lie at the same redshift as their central galaxies at projected distances of 35 and 151~kpc, respectively. We combine the spectroscopy with Tractor modeling from the Legacy Surveys and Prospector fits to the $griz$+W1/W2 photometry.

The distances of the three central host galaxies rule out a Galactic origin, and are consistent with the enormous intrinsic ORC diameters (300--500~kpc) based on photometry \citep{2022MNRAS.513.1300N}. Combining the optical and radio data, we find that the central hosts are massive, red ellipticals with little or no ongoing star formation. They host low-luminosity, radio-quiet AGN with low Eddington ratios. They are optical LINERs whose line emission is likely powered by shocks in ORC~4c, and which may be shock- or AGN-powered in ORCs 1c and 5c.

Previously, we argued based on optical IFS that the radio and ionized gas properties of ORC4c are consistent with a blast wave arising from a massive star-forming event in the past Gyr \citep{2024Natur.625..459C}. The current data support this view. The similarities of ORC1c, 4c, and 5c also support a common origin for ORCs 1, 4, and 5. The ionized gas properties of ORC1c and 5c are closer to those of massive ellipticals than those of ORC4c. ORCs 1c and 5c have \ha\ luminosities about 10$\times$ lower than in ORC4c (but still 10$\times$ higher than comparably massive ellipticals in the local Universe). Unlike ORC4c, in these two galaxies there is no evidence of a starburst event in the last Gyr. Thus, if they share a common origin with ORC 4, ORCs 1 and 5 may be due to an older event in which the resulting ionized gas shock emission has subsided, and in which the propagation speed of the forward shock was slower than in ORC4.

ORCs 2 and 3 likely have a different origin than the ORCs with prominent central galaxies. We find that the galaxy between the two ORCs, ORC23c, is a Type 2 AGN in a massive, red, disk galaxy. This supports the possibility that this double ORC is in fact a double-lobed radio galaxy.

Better constraints on the ionized gas properties of these systems, including possible extended emission requires sensitive and wide-field integral-field spectroscopy. KCWI observations of ORC~4 revealed extended, high-dispersion \ot\ emission that advance the blastwave model \citep{2024Natur.625..459C}. Deeper, spatially-resolved spectra would help distinguish between shock and AGN excitation in ORCs 1 and 4. They could also constrain the nature of ORCs 2 and 3 by probing other, potentially related, galaxies in the field and searching for extended line emission.

\section*{Acknowledgments}

The authors thank the Spring 2022 Observational Astronomy class at Rhodes College for workshopping the proposal that led to these observations.

Based on observations obtained at the international Gemini Observatory under Program IDs GN-2022A-FT-108 and GS-2022A-FT-110. Gemini is a program of NSF’s NOIRLab, which is managed by the Association of Universities for Research in Astronomy (AURA) under a cooperative agreement with the National Science Foundation on behalf of the Gemini Observatory partnership: the National Science Foundation (United States), National Research Council (Canada), Agencia Nacional de Investigaci\'{o}n y Desarrollo (Chile), Ministerio de Ciencia, Tecnolog\'{i}a e Innovaci\'{o}n (Argentina), Minist\'{e}rio da Ci\^{e}ncia, Tecnologia, Inova\c{c}\~{o}es e Comunica\c{c}\~{o}es (Brazil), and Korea Astronomy and Space Science Institute (Republic of Korea). The data was downloaded from the Gemini Observatory Archive at NSF’s NOIRLab.

This work was enabled by observations made from the Gemini North telescope, located within the Maunakea Science Reserve and adjacent to the summit of Maunakea. We are grateful for the privilege of observing the Universe from a place that is unique in both its astronomical quality and its cultural significance.

The Legacy Surveys consist of three individual and complementary projects: the Dark Energy Camera Legacy Survey (DECaLS; Proposal ID \#2014B-0404; PIs: David Schlegel and Arjun Dey), the Beijing-Arizona Sky Survey (BASS; NOAO Prop. ID \#2015A-0801; PIs: Zhou Xu and Xiaohui Fan), and the Mayall z-band Legacy Survey (MzLS; Prop. ID \#2016A-0453; PI: Arjun Dey). DECaLS, BASS and MzLS together include data obtained, respectively, at the Blanco telescope, Cerro Tololo Inter-American Observatory, NSF’s NOIRLab; the Bok telescope, Steward Observatory, University of Arizona; and the Mayall telescope, Kitt Peak National Observatory, NOIRLab. Pipeline processing and analyses of the data were supported by NOIRLab and the Lawrence Berkeley National Laboratory (LBNL). The Legacy Surveys project is honored to be permitted to conduct astronomical research on Iolkam Du’ag (Kitt Peak), a mountain with particular significance to the Tohono O’odham Nation.

%NOIRLab is operated by the Association of Universities for Research in Astronomy (AURA) under a cooperative agreement with the National Science Foundation.
LBNL is managed by the Regents of the University of California under contract to the U.S. Department of Energy.

This project used data obtained with the Dark Energy Camera (DECam), which was constructed by the Dark Energy Survey (DES) collaboration. Funding for the DES Projects has been provided by the U.S. Department of Energy, the U.S. National Science Foundation, the Ministry of Science and Education of Spain, the Science and Technology Facilities Council of the United Kingdom, the Higher Education Funding Council for England, the National Center for Supercomputing Applications at the University of Illinois at Urbana-Champaign, the Kavli Institute of Cosmological Physics at the University of Chicago, Center for Cosmology and Astro-Particle Physics at the Ohio State University, the Mitchell Institute for Fundamental Physics and Astronomy at Texas A\&M University, Financiadora de Estudos e Projetos, Fundacao Carlos Chagas Filho de Amparo, Financiadora de Estudos e Projetos, Fundacao Carlos Chagas Filho de Amparo a Pesquisa do Estado do Rio de Janeiro, Conselho Nacional de Desenvolvimento Cientifico e Tecnologico and the Ministerio da Ciencia, Tecnologia e Inovacao, the Deutsche Forschungsgemeinschaft and the Collaborating Institutions in the Dark Energy Survey. The Collaborating Institutions are Argonne National Laboratory, the University of California at Santa Cruz, the University of Cambridge, Centro de Investigaciones Energeticas, Medioambientales y Tecnologicas-Madrid, the University of Chicago, University College London, the DES-Brazil Consortium, the University of Edinburgh, the Eidgenossische Technische Hochschule (ETH) Zurich, Fermi National Accelerator Laboratory, the University of Illinois at Urbana-Champaign, the Institut de Ciencies de l’Espai (IEEC/CSIC), the Institut de Fisica d’Altes Energies, Lawrence Berkeley National Laboratory, the Ludwig Maximilians Universitat Munchen and the associated Excellence Cluster Universe, the University of Michigan, NSF’s NOIRLab, the University of Nottingham, the Ohio State University, the University of Pennsylvania, the University of Portsmouth, SLAC National Accelerator Laboratory, Stanford University, the University of Sussex, and Texas A\&M University.

BASS is a key project of the Telescope Access Program (TAP), which has been funded by the National Astronomical Observatories of China, the Chinese Academy of Sciences (the Strategic Priority Research Program “The Emergence of Cosmological Structures” Grant \#XDB09000000), and the Special Fund for Astronomy from the Ministry of Finance. The BASS is also supported by the External Cooperation Program of Chinese Academy of Sciences (Grant \#114A11KYSB20160057), and Chinese National Natural Science Foundation (Grant \#12120101003, \#11433005).

The Legacy Survey team makes use of data products from the Near-Earth Object Wide-field Infrared Survey Explorer (NEOWISE), which is a project of the Jet Propulsion Laboratory/California Institute of Technology. NEOWISE is funded by the National Aeronautics and Space Administration.

The Legacy Surveys imaging of the DESI footprint is supported by the Director, Office of Science, Office of High Energy Physics of the U.S. Department of Energy under Contract No. DE-AC02-05CH1123, by the National Energy Research Scientific Computing Center, a DOE Office of Science User Facility under the same contract; and by the U.S. National Science Foundation, Division of Astronomical Sciences under Contract No. AST-0950945 to NOAO.

\vspace{5mm}
\facilities{Gemini:Gillett (GMOS longslit), Gemini:South (GMOS longslit)}

\software{{\ttfamily extinction} (\url{https://extinction.readthedocs.io}), IFSFIT \citep{2014ascl.soft09005R,2017ApJ...850...40R}, {\ttfamily ppxf} \citep{2012ascl.soft10002C,2017MNRAS.466..798C}, {\ttfamily PypeIt} \citep{pypeit:joss_pub, j_xavier_prochaska_2023_7662288}, {\ttfamily pyphot} \citep{Fouesneau_pyphot_2022}, {\ttfamily speclite} (\url{http://dx.doi.org/10.5281/zenodo.8347108}), WebPlotDigitizer \citep{Rohatgi2022}}

%\appendix
\section{Appendix information} \label{sec:appendix}
%\restartappendixnumbering

\begin{figure*}
    \centering
    \includegraphics[width=\textwidth]{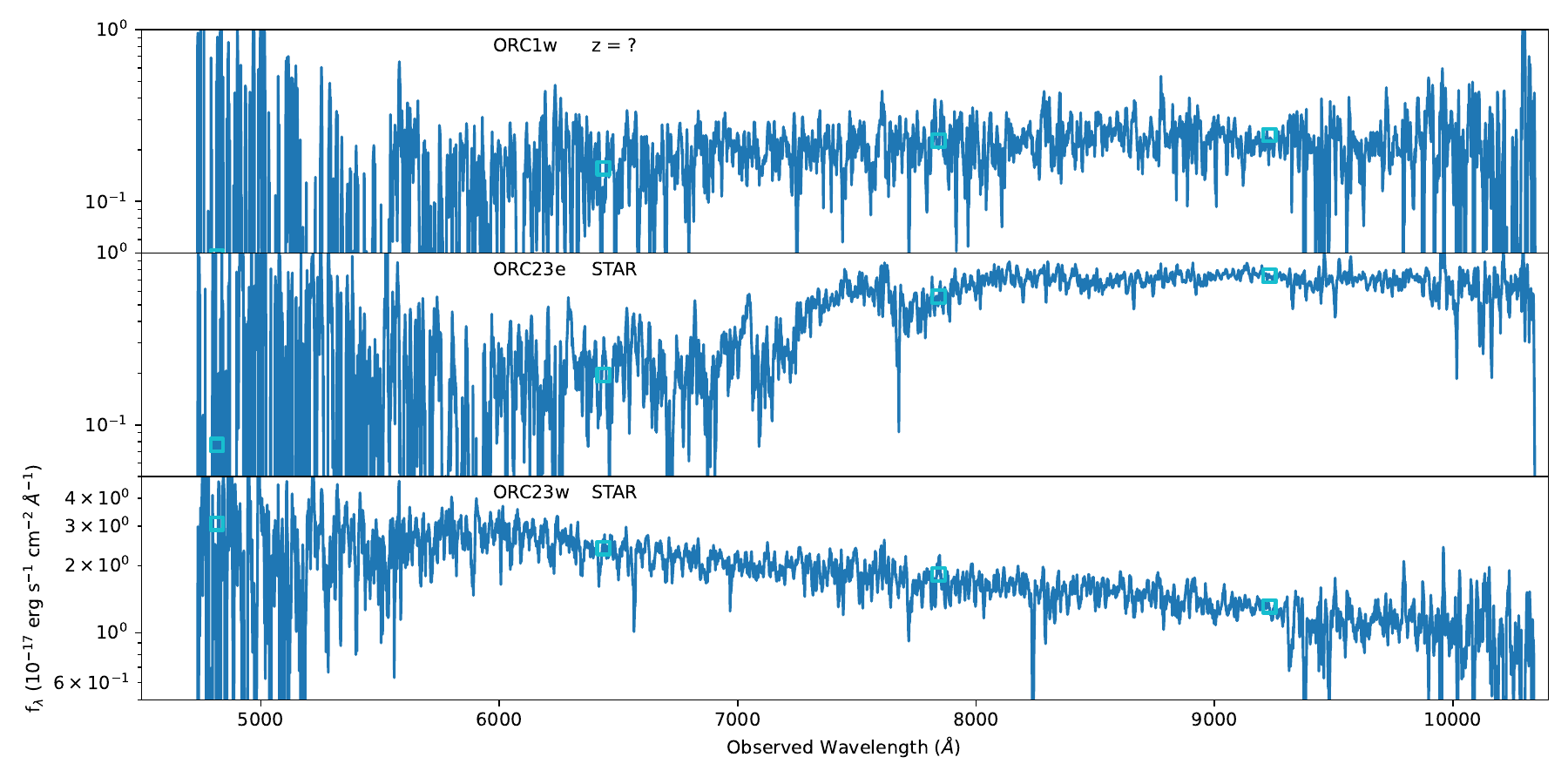}
    \caption{Observed-frame spectra of the unidentified source (ORC1w) and two stars (ORC23e/w).}
    \label{fig:spec-unided}
\end{figure*}

Here we display spectra of two stars and one un-IDed galaxy (Figure~\ref{fig:spec-unided}).

\bibliography{orc-spectra}{}
\bibliographystyle{aasjournal}

\end{document}